\documentclass[%
 reprint,
 amsmath,amssymb,
 aps,
]{revtex4-2}

\usepackage{graphicx}
\usepackage{dcolumn}
\usepackage{bm}
\usepackage{romannum}
\usepackage{amsmath}
\usepackage{graphicx}
\usepackage{xcolor}
\usepackage[caption=false]{subfig}
\usepackage{booktabs}   
\usepackage{relsize}

\begin{document}

\preprint{APS/123-QED}

\title{Detection of a universal core-halo transition in dwarf galaxies as predicted by Bose-Einstein dark matter}
\author{Alvaro Pozo$^{1,2}$, Tom Broadhurst$^{1,2,3}$, Ivan de Martino$^{4}$, Tzihong Chiueh$^{9,10}$,\\ George~F.~ Smoot$^{2,5,6,7,8}$, Silvia Bonoli,$^{2,3}$, Raul Angulo$^{2,3}$}

\affiliation{$^{1}$\textit{Department of Theoretical Physics, University of the Basque Country UPV/EHU, E-48080 Bilbao, Spain;\\ email:alvaro.pozolarrocha@bizkaia.eu; tom.j.broadhurst@gmail.com;}\\
$^{2}$\textit{DIPC, Basque Country UPV/EHU, E-48080 San Sebastian, Spain}\\
$^{3}$\textit{Ikerbasque, Basque Foundation for Science, E-48011 Bilbao, Spain}\\
$^{4}$\textit{Universidad de Salamanca,Facultad de Ciencias.,F\'isica Te\'orica, Salamanca, Plaza de la Merced s/n. 37008, Spain; ivan.demartino@usal.es}\\
$^{5}$\textit{Institute for Advanced Study and Department of Physics, IAS TT \& WF Chao Foundation Professor, Hong Kong University of Science and Technology, Hong Kong}\\
$^{6}$\textit{Energetic Cosmos Laboratory, Nazarbayev University, Nursultan, Kazakhstan}\\
$^{7}$\textit{Physics Department, University of California at  Berkeley CA 94720 Emeritus}\\
$^{8}$\textit{Paris Centre for Cosmological Physics, APC, AstroParticule et Cosmologie, Universit\'{e} de Paris, CNRS/IN2P3, CEA/lrfu,Universit\'{e} Sorbonne Paris Cit\'{e}, 10, rue Alice Domon et Leonie Duquet,	75205 Paris CEDEX 13, France  Emeritus}\\
$^{9}$\textit{Department of Physics, National Taiwan University, Taipei 10617, Taiwan}\\
$^{10}$\textit{National Center for Theoretical Sciences, National Taiwan University, Taipei 10617, Taiwan}}

\date{\today}

\begin{abstract}
\textbf{Recent discoveries of large halos of stars and dark matter around some of the lowest mass galaxies defy expectations that dwarf galaxies should be small and dense. Here we find large halos are a general feature of the well known dwarfs orbiting the Milky Way and also for the isolated dwarfs in the Local Group. Furthermore, these halos are seen to surround a dense core within each dwarf, with a clear density transition visible between the core and the halo at a radius of $\simeq 1.0{\rm kpc}$. This common core-halo structure is hard to understand for standard heavy particle dark matter where featureless, concentrated profiles are predicted, whereas dark matter as a Bose-Einstein condensate, $\psi$DM, naturally accounts for the observed profiles, predicting a dense soliton core in every galaxy surrounded by a tenuous halo of interfering waves. We show that the stellar profiles are accurately fitted by the core-halo structure of $\psi$DM, with only one fixed parameter, the boson mass. We also find independent consistency with the stellar velocity dispersions measured in these dwarf galaxies, which peak at the core radius and fall in the halo, at a level consistent with a boson mass of $\simeq 1.5\times 10^{-22}$ eV, based on independent dynamical work. Hence, dark matter comprised of light bosons, such as the axions generic in String Theory, provides a compelling solution for the structure of dwarf galaxies with stars that simply trace the dark matter profile of a Bose-Einstein condensate.} \\

\end{abstract}

\maketitle

\section{Introduction}

\begin{figure*}[htp]
	\centering
	\includegraphics[width=2.0\columnwidth,height=17cm]{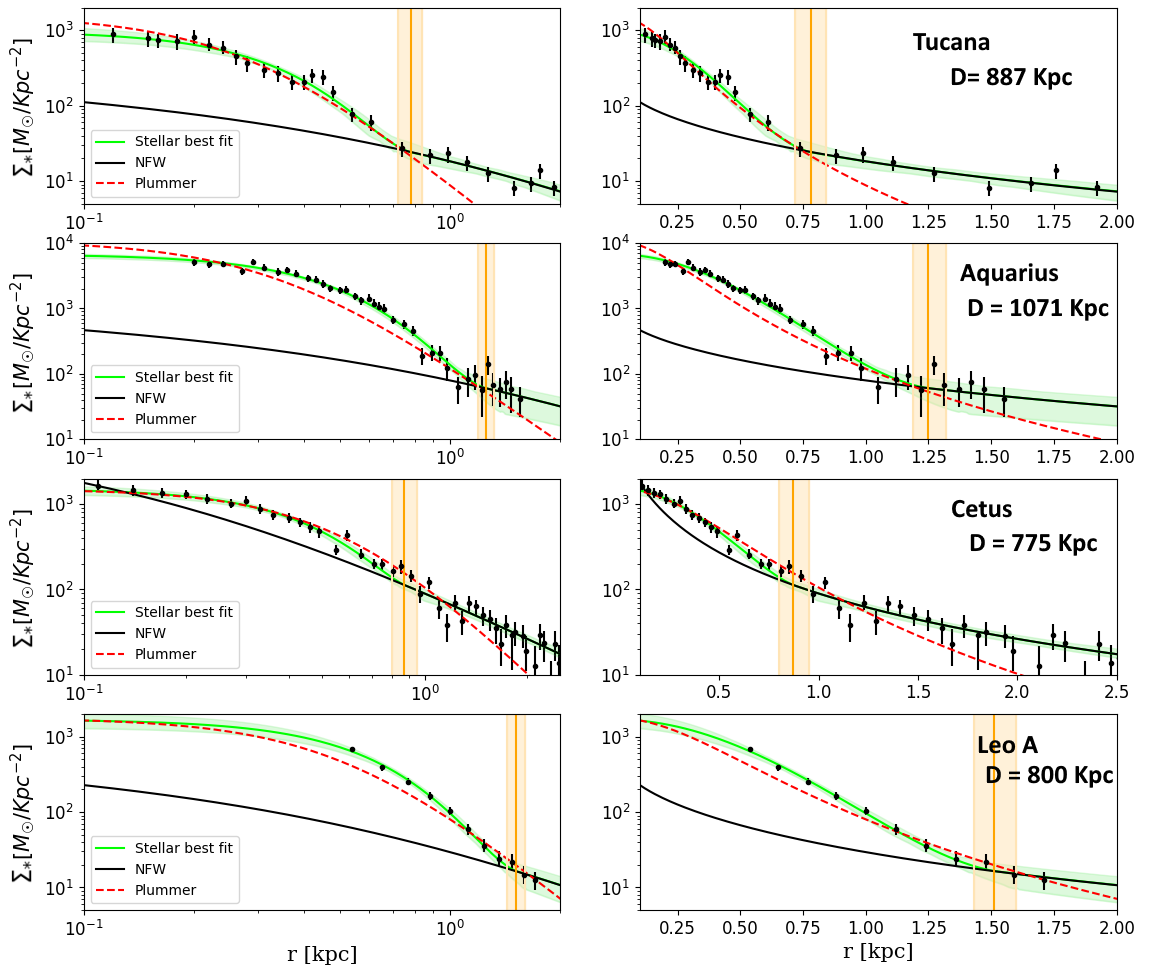}
	\caption{{\bf Isolated Dwarf Galaxies:} This figure shows the star count profiles versus dwarf galaxy radius for the well studied ``isolated" dwarf galaxies in the local group, lying outside the virial radius of the Milky Way. (In the right panels {\b D} shows the distance from the Milky Way galaxy center.)  Each dwarf galaxy has an extended halo of stars stretching to $\simeq 2$ kpc and most evident on the linear scale of right hand panel. Cores are also evident on a scale $<1$ kpc in each dwarf. A standard Plummer profile (red curve) is seen to fit approximately the core region but falls well short at large radius. Our predictions for light boson dark matter, $\psi$DM, are shown in green, where the distinctive soliton profile provides and excellent fit to the observed cores with the surrounded halo of excited states that average azimuthally to an approximately NFW-like profile beyond the soliton radius. The observed cores are excellent agreement with the predicted soliton, best seen on a log scale in the left panels,
	and the predicted $\psi$DM halo (grey curve) is also seen to match well the observed halos, including the characteristic density drop of about a factor of $\simeq 30$ predicted by $\psi$DM between the prominent core and tenuous halo at a radius $\simeq 1$ kpc indicated by vertical orange band. The best fit MCMC profile parameters are tabulated in the supplement. References for the data are: Tucana \cite{Gregory:2019}, Cetus\cite{McConnachie:2006}, Leo A\cite{Kang:2019} , Aquarius\cite{McConnachie:20062}}\label{fig1}
\end{figure*}

The origin of Dark Matter (DM) extends beyond the standard model of particle physics, representing a substantial portion of the cosmological mass density \citep{Cyburt:2016,Planck:2016}. DM is non-relativistic and primarily interacts through gravity, as evidenced by observations such as the Cosmic Microwave Background and galaxy power spectrum \citep{Markevitch:2004,Clowe:2006}. Cold Dark Matter (CDM) has traditionally been associated with heavy stable particles, yet the absence of new particle signatures in laboratories and inconsistencies with dwarf galaxy properties have fueled interest in alternative models \cite{Aprile:2018, Moore:1994,deBlok:2010,Marsh:2014,Klypin:1999, Safarzadeh:2021}.
One such model is Fuzzy Dark Matter (FDM) based on ultra-light axions, existing in a Bose-Einstein condensate state with wave-like structures\citep{Schive:2014,Schive:20142,Schive:2016}. FDM addresses CDM's challenges in resolving galactic-scale behavior, offering predictions like the formation of solitonic standing waves matching known theoretical solutions. These solitons exhibit flat-cored density profiles and follow scaling relations with host virial masses, leading to more compact solitons in more massive galaxies \citep{Schive:20142}. At scales larger than the de Broglie scale, FDM simulations align with CDM simulations, ensuring consistency with established observations of large-scale structure and the Cosmic Microwave Background. Despite contrasting with heavy fermions from supersymmetry, light bosons in FDM provide a viable non-relativistic explanation for dark matter's observed coldness.

Galaxies with low velocity dispersions $\lesssim 15~{\rm km/s}$, indicating small masses, are classed as ``dwarfs" with half of the stars detected typically within only $r_{1/2}\simeq 0.3$ kpc. So it is surprising that several low mass dwarfs are now known to possess large halos of stars and dark matter extending to over several kpc, defying the dwarf definition. This includes two spectroscopically detected halos that are dynamically dominated by dark matter around the Tucana II and AndXXI dwarf spheroidal galaxies \cite{Chiti:2021,Collins:2021}. This adds to the case of Crater II, a dwarf that extends to over 3 kpc despite its very low velocity dispersion of $\simeq$ 3 km/s and also the ``ghostly" Antlia II of extremely low surface brightness extending over 4 kpc with a dispersion of 6 km/s, discovered serendipitously using GAIA satellite proper motions \cite{Torrealba:2016,Collins:2021}. Such ``large dwarfs" are at odds with the compact, high concentration profiles predicted for low mass galaxies in N-body simulations of standard heavy particle dark matter of cold Dark Matter (CDM), where dwarfs are predicted to have the highest internal density of dark matter of any galaxy, reflecting the relatively high Universal mean density at earlier times when dwarf galaxies were first formed. In this CDM context, the extended halos of dwarfs have been qualitatively attributed to tidal effects induced by the Milky Way, or Andromeda \cite{Torrealba:2019, Collins:2021, Chiti:2021}, based on simulations that show stars may be periodically stripped or shocked near pericenter to beyond the tidal radius, generating halo-like extensions of enhanced velocity dispersion. In contrast, the outer velocity dispersions observed in most dwarf halos appear to be significantly lower than in their cores \cite{Wilkinson:2004, Fabrizio:2016, Collins:2021}. Strong tidal effects are expected for only a minority of orbiting dwarfs on eccentric orbits with small pericenters and so it is important to examine the generality of stellar halos to determine whether such halos are atypical or perhaps a common structural component of dwarf galaxies. Two such cases of the Milky Way dwarfs are definitively established to be in the process of being tidally stripped, namely the Sagittarius and Tucana III dwarfs which show opposing pairs of tidal arms \cite{Li:2018,Newby:2013}, representing only $\simeq 5\%$ of the dwarfs orbiting the Milky Way and both of these dwarfs have relatively small orbits.


\section{The Wave Dark Matter Halo}

Ultralight bosons, such as Axions, explored in relation dark matter \cite{Widrow:1993,Hu:2000} in their simplest version, without self-interaction, the boson mass is the only free parameter, which if sufficiently light means the de-Broglie wavelength exceeds the mean free path set by the density of dark matter, so these bosons can satisfy the ground state condition for a Bose-Einstein condensate described by the coupled Schroedinger-Poisson equation, that in comoving coordinates reads:
\begin{align}
& \biggl[i\frac{\partial}{\partial \tau} + \frac{\nabla^2}{2} - aV\biggr]\psi=0\,,\\
& \nabla^2 V =4\pi(|\psi|^2-1)\,.
\end{align}
Here $\psi$ is the wave function, $V$ is the gravitation potential and $a$ is the cosmological scale factor. The system is normalized to the time scale $d\tau=\chi^{1/2} a^{-2}dt$, and  to the scale length $ \xi = \chi^{1/4} (m_B/\hbar)^{1/2} {\mathbf x}$, where $\chi=\frac{3}{2}H_0^2 \Omega_0$  where $\Omega_0$ is the current density parameter \cite{Widrow:1993}. 

Recently, it has proved possible with advanced GPU computing to make reliable, high dynamic range cosmological simulations that solve the above equations, \cite{Schive:2014,Schwabe:2016,Mocz:2017,May:2021} that evolve to produce large scale structure indistinguishable from CDM, but with virialized halos characterized by a solitonic core in the ground state that naturally explains the dark matter dominated cores of dwarf spheroidal galaxies \cite{Schive:20142}. Another, important feature arising from simulations is that the central soliton is surrounded by an extended halo with a ``granular" texture on the de-Broglie scale, due to interference of excited states, but which when azimuthally averaged follows closely the Navarro-Frank-White (NFW) density profile \cite{Navarro:1996,Woo:2009,Schive:2014,Schive:20142}.

The fitting formula for the density profile of the solitonic core in a $\psi$DM halo  is obtained from cosmological simulations \cite{Schive:2014,Schive:20142}:
\begin{equation}\label{eq:sol_density}
\rho_c(r) \sim \frac{1.9~a^{-1}(m_\psi/10^{-23}~{\rm eV})^{-2}(r_c/{\rm kpc})^{-4}}{[1+9.1\times10^{-2}(r/r_c)^2]^8}~M_\odot {\rm pc}^{-3}.
\end{equation}
where the values of the constants are: $c_1=1.9$, $c_2=10^{-23}$ , $c_3=9.1\times10^{-2}$; $m_\psi$ is the boson mass
and $r_c$ is the solitonic core radius. The latter scales with the product of the galaxy mass and boson mass, obeying the following the scaling relation which has been derived from our simulations \cite{Schive:20142}:
\begin{equation}\label{eq:sol_radius}
r_c=1.6\biggl(\frac{10^{-22}}{m_\psi}  eV \biggr)a^{1/2}
\biggl(\frac{\zeta(z)}{\zeta(0)}\biggr)^{-1/6}
\biggl(\frac{M_h}{10^9M_\odot}\biggr)^{-1/3}kpc
\end{equation}

Where $a=1/(1+z)$ and z refers to the Redshift ( see \citep{Bryan:1998,Schive:20142} for original definitios). Beyond the soliton, at radii larger than a transition scale ($r_t$), the simulations also reveal the halo is approximately NFW in form, presumably reflecting the non-relativistic nature of condensates beyond the de Broglie 
scale, and therefore the total density profile can be written as:
\begin{equation}\label{eq:dm_density}
\rho_{DM}(r) =
\begin{cases} 
\rho_c(r)  & \text{if \quad}  r< r_t, \\
\frac{\rho_0}{\frac{r}{r_s}\bigl(1+\frac{r}{r_s}\bigr)^2} & \text{otherwise,}
\end{cases}
\end{equation} 
and:

 In detail, the scale radius of the solitonic solution, which represents the ground state of the Schrodinger-Poisson equation, is related to size to the halo through the uncertainty principle. From cosmological simulations, the latter is found to hold non-locally, relating a local property with a global one (for more details we refer to \cite{Schive:20142}).

\section{Dynamical Model of Galaxies in Wave Dark Matter Halo}

 The classical dwarf galaxies are known to be dominated by DM, and so the stars are treated as tracer particles \cite{Gregory:2019,McConnachie:2006,McConnachie:20062,Kang:2019} moving in the gravitational potential generated by DM halo density distribution.

In this context, the corresponding velocity dispersion profile can be predicted by solving the spherically symmetric Jeans equation:
\begin{equation}\label{eq:sol_Jeans}
\frac{d(\rho_*(r)\sigma_r^2(r))}{dr} = -\rho_*(r)\frac{GM_{DM}(r)}{r^2}-2\beta\frac{\rho_*(r)\sigma_r^2(r)}{r},
\end{equation}
where $M_{DM}(r)$ is the mass  DM halo obtained by integrating the spherically symmetric density profile in Eq. \eqref{eq:dm_density}, $\beta$ is the anisotropy parameter (see Binney \& Tremaine 2008\cite{Binney:2008}, Equation (4.61)), and $\rho_*(r)$ is the stellar density distribution defined by the solitons wave dark matter imprint:
\begin{equation}\label{eq:stellar_density}
\rho_{*}(r) =
\begin{cases} 
\rho_{1*}(r)  & \text{if \quad}  r< r_t, \\
\frac{\rho_{02*}}{\frac{r}{r_{s*}}\bigl(1+\frac{r}{r_{s*}}\bigr)^2} & \text{otherwise,}
\end{cases}
\end{equation}
where
\begin{equation}\label{eq:stellar_1}
\rho_{1*}(r) = \frac{\rho_{0*}}{[1+9.1\times10^{-2}(r/r_c)^2]^8}~N_* {\rm kpc}^{-3}
\end{equation}

Here, $r_{s*}$ is the 3D scale radius of the stellar halo corresponding to $\rho_{0*}$ the central stellar density, $\rho_{02*}$ is the normalization of $\rho_{0*}$ at the transition radius and the transition radius, $r_t$, is the point where the soliton structure ends and the halo begins at the juncture of the core and halo profiles. 

Finally, the predicted  velocity dispersion profile can be projected along the line of sight to compared with the observations, as presented in Figure~4:
\begin{equation}\label{eq:sol_projected}
\sigma^2_{los} (R) = \frac{2 }{\Sigma(R)} \int_{R}^{\infty} \biggl(1-\beta \frac{R^2}{r^2}\biggr) \frac{\sigma^2_r(r)\rho_*(r)}{(r^2-R^2)^{1/2}} r dr\,\,
\end{equation}
where 
\begin{equation}
\Sigma(R) =2\int_{R}^{\infty} \rho_*(r)(r^2-R^2)^{-1/2}rdr\,. 
\end{equation}

\section{Results}

 Here we examine the outer profiles of all known classical dwarf spheroidal galaxies (dSph) in the local neighbourhood, where stars can be individually counted to large radius, so the entire stellar profile can be traced free of surface brightness limitations. We start with the best studied  ``isolated" dwarfs lying beyond the virial radius of the Milky Way, and understood not to have interacted tidally with the major members of the Local group \cite{Gregory:2019,Fraternali:2009,Taibi:2018,Kirby:2017}. These isolated dwarfs include  Cetus, Tucana, Aquarius and Leo A, which have small velocity dispersions $\simeq 10~{\rm km/s}$ and old spheroidal stellar populations. Their stellar profiles are shown in Figure~1, where their large radial extents are visible to over $\simeq 2$ kpc \cite{Gregory:2019,McConnachie:2006,McConnachie:20062} (Note we have adopted the independent star counts directly as published, without rebinning, including many independent teams, that we reference in Table~1). It is also apparent from Figure~1 that the stellar halos of these dwarfs extend radially from a well defined core, with a clear transition in density between the core and the halo. The core is reasonably well fitted by the standard Plummer profile in each dwarf (red curve in Figure 1), but falls well short in the halo region beyond a transition radius indicated in Figure~1 (vertical orange band), so it is clear from Figure 1 that alone, neither the Plummer not the NFW profile fits the full stellar profiles of these dSph galaxies, as unlike the data, the Plummer profile falls well short in the halo and the NFW profile has no core.


\begin{figure*}[htp]
	\centering
	\includegraphics[width=2\columnwidth,height=9cm]{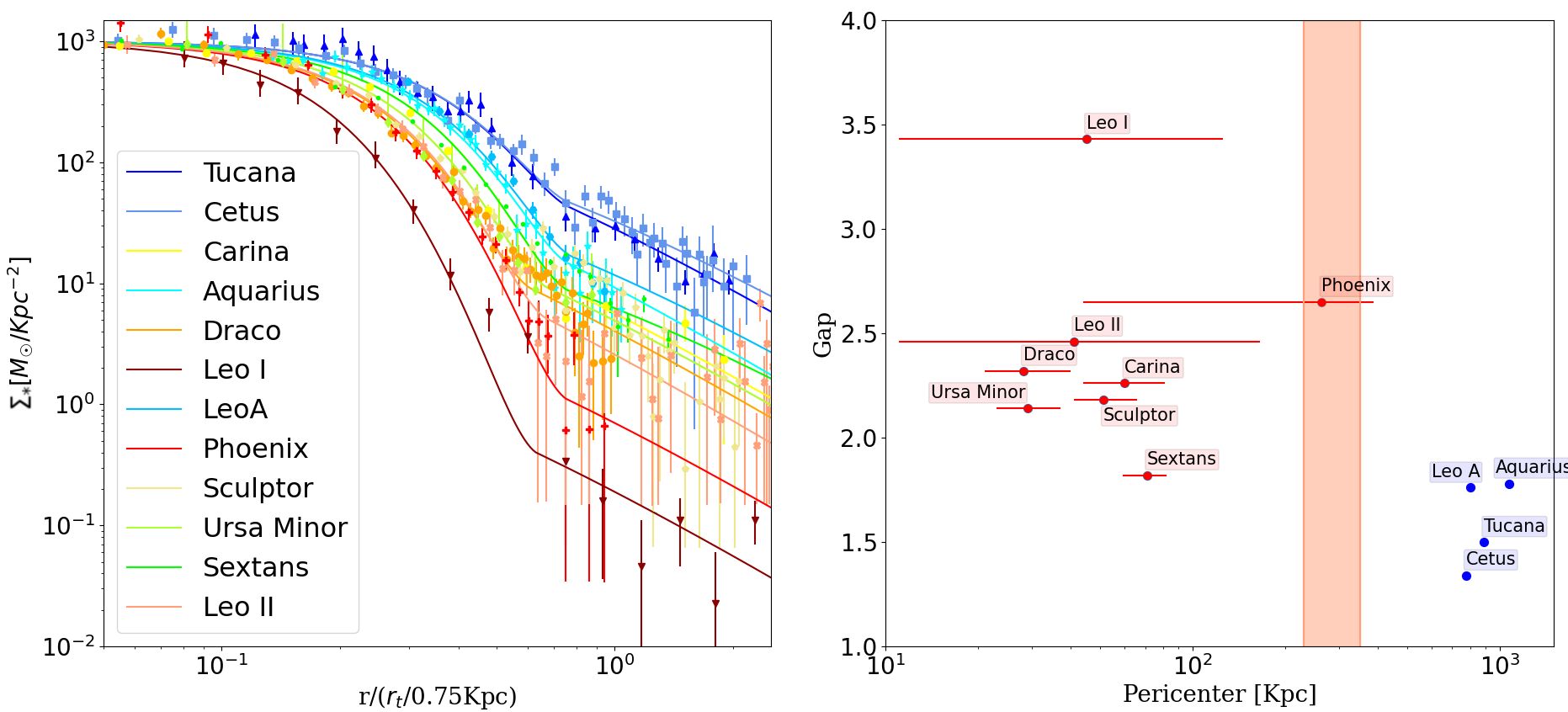}
	\caption{Stellar profiles of classical dwarfs orbiting the Milky Way and also the four well known ``isolated" dwarfs that lie beyond in the Local Group (all listed in Table 1), and rescaled by their measured transition radius, revealing these profiles have a common core-halo form that is more pronounced for the orbiting dwarfs (redder colours) for which the halo density is generally lower than the ``isolated" dwarfs (bluer profiles). {\bf The right hand panel} compares the change in stellar density, or ``gap", between the core and halo against the pericenter radius estimated from GAIA proper motions, showing the isolated dwarfs beyond the Milky Way (blue points) have generally smaller transition amplitudes compared to most of the orbiting dwarfs (red points).}\label{fig2}
\end{figure*}

\begin{figure*}[htp]
	\centering
	\includegraphics[width=2\columnwidth,height=9cm]{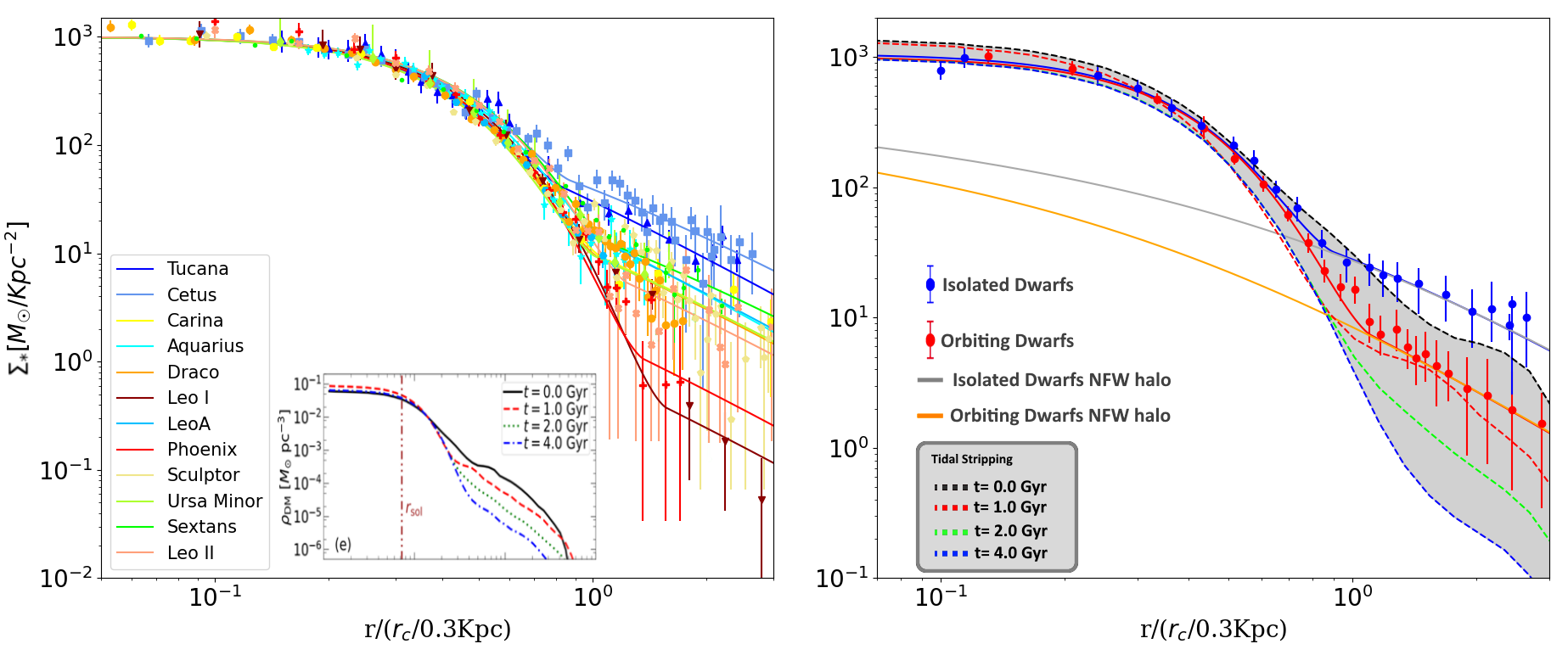}
	\caption{{\bf The left hand panel} compares all the dwarf profiles scaled ($\Sigma_{0*}$=1000 $M_\odot/Kpc^{-2}$) and all the core radii set to 0.3 kpc. The profiles are very similar in the core region, differing in the extended halo region  relative to the core, with the isolated galaxies (bluer colours), having denser halos than the orbiting dwarfs (redder colours). {\bf The right hand panel} shows the projected  simulation profiles for $\psi$DM by Schive, Chiueh \& Broadhurst (2020) \cite{Schive:2020}, where little dependence on the degree of tidal stripping is predicted for the core region, in contrast to the halo region bracketed in grey where stripping is significant. The duration of stripping is indicated by the legend and also shown in the inset of the left hand panel  spanning 1-4Gyrs, and matching well the observed range of halo profiles. Note, the simulations predict the halo slope is relatively shallow and fairly independent of the degree of stripping, in good agreement with the mean halo profile of the isolated dwarfs (averages of the normalized profiles shown in the left panel, blue data) and the orbiting dwarfs (red data points) including the larger core-halo transition of the orbiting dwarfs. The NFW profile fits to the halos are also shown and can be seen to fall well below the prominent cores. These last two (organge and grey lines) are examples of the required pure NFW profiles that fit the observed stellar halo slopes of the isolated and orbiting dwarfs, respectively.}
	 \label{fig3}
\end{figure*}

\begin{figure*}[htp]
	\centering
	\includegraphics[width=2.00\columnwidth,height=18cm]{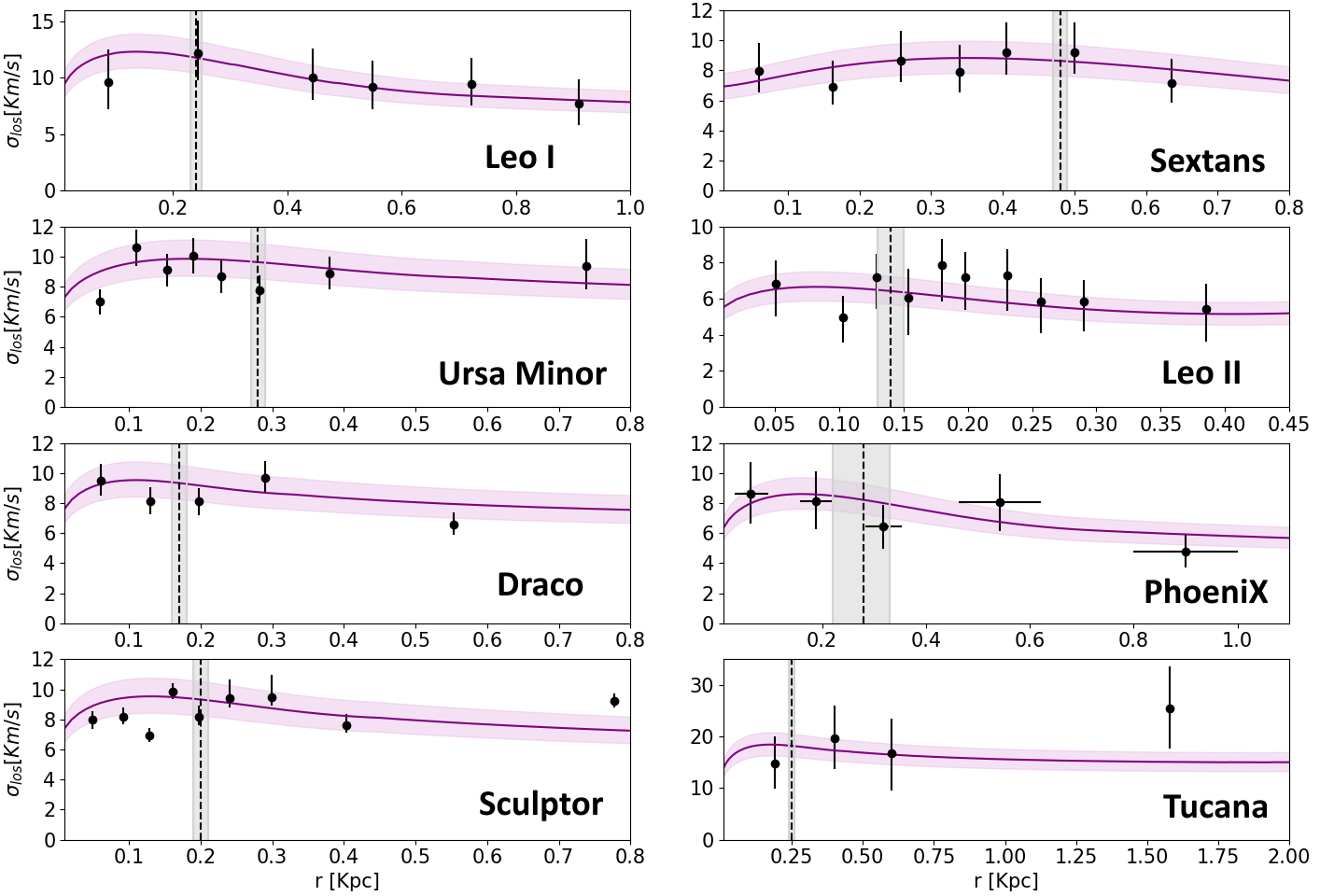}
	\caption{{Comparison of the well measured dispersion profiles (black data points) of the classical dwarfs, with the $\psi$DM model fit obtained from the stellar profiles with our MCMC analysis above (purple curves), indicating good consistency in general with the characteristic $\psi$DM dispersion profile that can be seen to peaks near the core radius in each case, indicated by the vertical band (listed in Table~1). The range of model profiles shown spans  the range of boson mass of $1.3-1.7\times10^{-22}$eV, and assumes a modest fixed anisotropy parameter, $\beta$ = -0.5}}\label{fig6}
\end{figure*}

We now turn to the well studied classical dwarf galaxies orbiting the Milky Way. For these, detections of stars at surprisingly large radius beyond the estimated tidal radius \cite{Irwin:1995} and originally termed ``..`extra tidal stars' - for convenience..." \cite{Irwin:1995}. Subsequently, such stars have been assumed to be
tidally stripped, motivated by simulations where temporary extensions can be generated for dwarfs on rather radial orbits. However, the predicted enhancement of the velocity dispersion for stripped in the simulations may conflict with the observations, as reduced velocity dispersion are found at large radius \cite{Wilkinson:2004}. In Figure~2 we compare the deepest, wide-field profiles for the classical dwarfs, for which a general core-halo structure is seen extending beyond $\simeq 1.0$ kpc, with very similar outer profile gradients that are relatively shallow and extend well beyond the standard Plummer profile. Distinct cores are also evident, with a clear density transition between the core to the halo visible in most cases, (individual fits shown in the Methods section, and fitted parameters listed in Table~1).  In Figure 3a we scale the stellar profiles by their individually fitted core radius, which reveals more clearly the distinctive core-halo profile, with relatively larger density transitions seen for these orbiting dwarfs (red data, Figure 3b) than the mean isolated dwarf profile (blue data, Figure 3b). The red  "orbiting" data and the blue  "isolated" data of Figures 3b have been computed after calculating the means of the respective scaled profile of Figure 3a (see Figure 2b to visualize the members of each group).

The observed profiles are seen to be very similar within the core region, below the mean transition radius of 0.75kpc, whereas at larger radius the profile varies by two orders of magnitude, but with similar gradients. 

We define a transition gap to quantify the change in density between the core and the halo; $\Delta_{C-H}= \log{\rho_{C}/\rho_{H}}$, where $\rho_{C}$ is the asymptotic central core stellar density, and shown in Figure~2b. The value of the gap is generally larger for the orbiting dwarfs than the isolated dwarfs, with $\Delta_{C-H} > 2.0$, see Figure~2b. The largest core-halo transition is found for Leo I, with $\Delta_{C-H}\simeq 3.5$ (Figure ~2b) and so it is interesting that although Leo I is relatively distant now at 250 kpc, its orbital pericenter is now established by the GAIA satellite to be only $\simeq$ 50 kpc, moving on an eccentric orbit (see table 1) where tidal stripping is enhanced during close approaches \cite{Sohn:2007}. Tidal effects are also claimed for Ursa Minor \cite{Martinez-Delgado:2001}, Sculptor \cite{Walker:2011,Amorisco:2012} and Carina dwarfs\cite{Battaglia:20122}.

The core-halo structure we have uncovered here is a general feature of these classical dwarfs and far from expected for standard CDM, where low mass galaxies should be concentrated and core-less, with no inherent density transition. A fair fit to the inner region is provided by the Plummer profile, standard in Jeans analysis, but clearly does not extend into the halos. Expectations of small sizes and high concentrations for low mass galaxies are conditioned by CDM simulations, however a physically very different explanation for dark matter as light bosons is now understood to naturally form wide cores and extended diffuse halos. This is seen in the first simulations of dark matter (DM) as a Bose-Einstein condensate \cite{Schive:2014, Schive:20142, Schwabe:2016} revealing an unanticipated, rich wave-like structure on the de Broglie scale, described simply by a coupled Schrodinger-Poisson equation for the mean-field behaviour under self-gravity, hence the term $\psi$DM. Condensates are inherently non-relativistic; hence, $\psi$DM behaves as "cold" dark matter on large scales, exceeding the de Broglie wavelength, which is statistically indistinguishable from CDM, as demonstrated in the first simulations \cite{Schive:2014}. Several unique predictions are now established for $\psi$DM, including a dark core within each galaxy that follows the soliton solution of the Schrodinger-Poisson equation (see methods) with a radius, $r_{sol}$, set by the de Broglie wavelength. Implying also that this core radius should be largest in lower mass galaxies, $m_{gal}$, of lower momentum, scaling as:  $r_{sol}\propto m_{gal}^{-1/3}$ as predicted by Schive et al. (2014b) \cite{Schive:20142} and verified in independent simulations, \cite{Schwabe:2016, Niemeyer:2020,Mocz:2017, Veltmaat:2019,Hui:2020}.


The predicted soliton profile has been shown to match well the Fornax dwarf, for which the dynamical data extends beyond the core radius allowing a determination of the boson mass of $m_{\psi} \simeq 10^{-22}{\rm eV}$ \cite{Schive:2014} and this supported by Jeans analyses of other classical dwarfs \cite{Chen:2017}. This result seems to be inconsistent with the bound given by \cite{GonzalesMorales:2017}, who required a boson mass $<$ 0.4$\times10^{-22}$eV to fit Fornax's and Sculptor's cores,  result of probably overestimated cores bigger than 1kpc, that are in clear disagreement with the extracted sizes $<$ 0.5kpc from the stellar profiles of all dwarfs. The smaller the boson mass the higher the effective cut of mass for $\psi$DM, which depends on the detailed merger history with simulations now estimating the effect at the level. A useful new reference that deliberately simulates the satellite population predicted for a Milky Way type massive galaxy by \cite{Nori:2023}  finds agreement in the number of satellites for a light boson mass, (similar to our adopted value), so that is reassuring, in particular, they state:  “With the relatively low FDM mass, $ 1.5\times10^{-22}$eV adopted in this work – chosen to highlight the effect of FDM dynamics on galaxy formation processes…the number of dark matter satellites in FDM is approximately the same of the luminous satellites of a Milky-way like galaxy". Assuming that such an FDM model is valid would require almost all of the dark satellites to have a luminous component, which is not favoured for CDM based models of galaxy formation where feedback is reasonably invoked at low mass.


Here we examine the unique prediction that the density of $\psi$DM should transition sharply between the soliton core and the surrounding halo, as the soliton forms a prominent core that contrasts by over an order of magnitude in density above the halo. This transition is predicted to be distinct even though observations are made in projection, because the soliton core is close to a Gaussian and hence its sharp 3D boundary is preserved in 2D at the core radius. It is important to appreciate this prominent core is quite unlike the behaviour of smooth cores employed, where the core is continuous in density with the halo. In contrast, the $\psi$DM core is a stable standing wave that is a gravitationally self-reinforcing \cite{Schive:20142} with a pronounced overdensity predicted to be about $> 30$ times denser than the surrounding halo in the case of low mass galaxies relevant here, of $\simeq 10^{10} M_\odot$  (see Figures 1 \& 2 of Schive et al. 2014b \cite{Schive:20142}). It is also clear that the soliton core is relatively stable to tidal stripping compared to the halo, as shown in recent simulations of dwarf galaxies orbiting the Milky Way \cite{Schive:2020} as the soliton by nature is self reinforcing. The halo comprises excited de Broglie scale waves that fully modulate the density but is seen in the simulations to average azimuthally to an approximately Navarro-Frenk-White (NFW), reflecting the cold, non-relativistic nature of the condensate on scales exceeding the de Broglie wavelength \cite{Schive:2016}. 

We first perform MCMC based $\psi$DM profile fits to the isolated dwarfs, shown in Figure 1. The shape of the soliton core profile is fully characterised by the core radius, $r_c$. For the halo we fit an NFW profile with scale radius $r_{s*}$ and normalization $\rho_{o*}$. The only other free parameter we require is the transition radius, $r_t$, defining the radius of the density transition between the soliton and NFW profile which we vary within a prior range indicated by the $\psi$DM simulations, of $2-4\times r_c$ (see Methods section), with the best parameters and uncertainties listed in Table~1 and in Figures 6-9 in the Supplement. Note, we have adopted the independent star counts as published, without rebinning, from many independent teams, that we reference in the corresponding captions and Table~1. Moreover, it is important to point out that the profile fit parameters in Table 1 are independent of the choice of boson mass we have adopted, $1.5\times10^{-22}$ eV, as indicated by dynamical work \cite{Chen:2017}, which we find is consistent with our mean dispersion profile derived in Figure 4.

We now compare the $\psi$DM model with the orbiting dwarfs, with new $\psi$DM simulations that quantify the effect of tidal stripping of a dwarf galaxy orbiting within a Milky Way sized halo, shown in Figure 3b. The main effect of tidal stripping is to strip the relatively tenuous halo, thereby enhancing the density transition at the core-halo radius, providing a natural explanation for the generally larger "mass gap" of the orbiting dwarfs compared to the isolated dwarfs, plotted in Figure~2b. In detail, the family of profiles for the orbiting dwarfs is seen in Figure 3b to span the predicted range when tidal stripping is included, spanning several Gyrs, as can be seen in Figure 3a. The Sextans dwarf galaxy has a core-halo transition that is intermediate between orbiting and isolated dwarfs, with an extensive core-halo structure and a transition gap of $\Delta_{C-H} \simeq 1.8$ (Figure~2b), suggesting that Sextans is less tidally stripped than the other orbiting dwarfs, a conclusion supported by the undisturbed morphology and simple internal dynamics noted for Sextans \cite{Roderick:2016,Okamoto:2017}, indicating Sextans may have become bound to the Milky Way relatively recently.

We now make an independent, dynamical consistency check of our $\psi$DM fitted stellar profile fits
above, using the well resolved velocity dispersion profiles available for most of the classical dwarfs, by inputting our bestfit $\psi$DM density profile for each dwarfs into the Jeans equation (see methods section) to predict the corresponding
velocity dispersion profile for comparison with the data for each dwarf. The form of these predicted velocity dispersion profiles should peak just beyond the core radius and then decline into the lower density halo, as shown in Figure~4, where consistency is evident in each case, both in terms of the form of the profile and in terms of the stellar core radius (vertical lines in Figure~4). The amplitude of the dispersion profiles peaks at a mean level of $\simeq 10km/s$ and together with the mean core radius of $\simeq 0.3kpc$ provides an estimate of the boson mass of $1.3-1.7\times10^{-22}$eV (see Supplement) as indicated by the spread in the model curves in Figure~4. This is similar to the boson mass from other dynamical studies for these classical dwarfs for $\psi$DM \cite{Schive:2014,Chan:2020,Broadhurst:2020}, and for the intermediate mass galaxy DF44 \cite{Pozo:2020} and also consistent with the boson mass estimate for the claimed soliton core within the Milky Way of $\simeq 100$pc, which is smaller than for dwarf galaxies estimated here, reflecting the inverse momentum dependence of the de Broglie scale\cite{Schive:20142}.

In contrast, the gas-rich dwarf galaxies classified as Low Surface Brightness (LSB) galaxies favor larger cores of several kiloparsecs (Kpc) based on HI gas rotation fields, which are much larger than the ~0.3 Kpc cores we have derived here for the dSph stars. These wider HI cores for LSB galaxies have shallow linear profiles that rise more slowly with radius, as predicted for NFW dark matter halos. It has been claimed that these cores are in tension with standard Cold Dark Matter (CDM) models and are significantly wider than the cores predicted for $\psi$DM with a boson mass of approximately $10^{-22}$ eV \cite{Bar:2019}, which is the value favored by dSph dynamics. Identification of bars in the HI gas rotation fields and other non-circular motions, and to a lesser extent, gas pressure support, has been found to be typical in high-resolution HI rotation field data, leading to an underestimate of the circular Keplerian velocity. This complexity in the inner gas motions is seen in CDM-based simulations and has been shown to lead to a significant underestimation of the Keplerian rotation. When averaged azimuthally, this results in a shallow rotation curve that underestimates the central mass density profile, leading to the spurious conclusion of wide cores for these LSB galaxies, as outlined by Oman et al. \cite{Oman:2019}. We note that these LSB galaxies are more massive than the dSph class of galaxies studied here, extending to $10^{11} M_\odot$, and hence the soliton core radius predicted in the $\psi$DM context is generally several times smaller than for dSph galaxies, around 0.1 Kpc. This makes it difficult to resolve such a small core at the typical distances of these LSB galaxies on galactic scales, rendering the identification of any such small core infeasible in HI studies. Moreover, the core density vs radius relation has been explored for a general class of light particles by \cite{Deng:2018}, including both fermions and bosons where steep inverted relations are generally predicted, as with $\psi$DM and compared with the relatively large cores fitted to HI rotation curves of massive galaxies with gas and bar dynamical complications that are absent for the DM dominated dSph galaxies explored here and may be significantly pressure supported and/or broadened by non-circular gas velocities.

\section{ Discussion and Conclusions}

We conclude that the class of dSph galaxies have a common core-halo structure, with consistency found between the stellar profiles and dynamics in the context of $\psi$DM, implying stars trace well the dark matter in these DM dominated galaxies. It is now understood that the equilibrium structure of $\psi$DM halos slowly relaxes with wave interference in the halo by continuously scattering stars incorporated as test particles in the latest $\psi$DM simulations \cite{Schive:2020, Dutta:2021}. This effect has been argued can account for the "thick disk" of the Milky Way, for a light boson of $\simeq10^{-22}$eV\cite{Church:2019}, similar to the boson mass estimates for dSph cores. This dynamically based light boson mass has been claimed to be in conflict with indirect estimates from the Lyman-$\alpha$ power spectrum of $<10^{-21}eV$. This complaint relies on analogy with warm dark matter because hydrodynamical simulations have not been achieved for $\psi$DM. Furthermore, such Ly-forest based modelling does not include an AGN contribution to reionization, as argued by\cite{Madau:2015,Padmanabhan:2021} and reinforced by $z\simeq 6$ detections of double peaked Ly$\alpha$ emitters \cite{Hu:2016,Bosman:2020,Gronke:2020} and by wide ``gaps" of enhanced photoionisation in the forest at $z>5$ \cite{Becker:2015}. Such effects point to the possibility of sparsely distributed AGNs\cite{Gangolli:2021} enhancing the variance and hence the power spectrum amplitude above standard reionization predictions that are based on more smoothly distributed galaxy photoionization.

The relatively high bound of $m_b > 10^{-19}$ eV, recently claimed based on two extreme dwarf galaxies, Segue I $\&$ II \cite{Dalal:2022}, relies mainly on the viability of the non-detection of internal velocity dispersion of Segue II, rather than on an open star cluster as may be the case. On the other hand, Segue I is certainly a bona-fide Ultra-faint dark matter dominated dwarf with a well-detected velocity dispersion of 5.4 km/s. A Jeans analysis of this dwarf prefers a relatively high boson mass of $10^{21.5}$ eV \citep{Pozo:2024}. This is an order of magnitude difference. Indeed, the UFD class as a whole fits well with this higher boson mass. A case for two boson species has been made to encompass both UFD and dSPh classes by some of us \cite{Pozo:2024} and is motivated by the Axiverse scenario, generic in String Theory and hence consistent with \cite{Dalal:2022}, in terms of the established UFD galaxy Segue I.

We can look forward to direct lensing tests of $\psi$DM that are predicted to be sensitive to the pervasive de Broglie scale perturbations within lensing galaxy halos, measurably affecting image magnifications and locations around the Einstein ring\cite{Chan:2020,Hui:2020} as recently claimed by \cite{Amruth:2023} or by direct detection of the Compton wavelength imprinted on pulsar timing \cite{deMartino:2017,Luu:2024}. It is also possible that oscillation modes of the soliton, generated by the unceasing wave motion in the halo, may result in a random walk of centrally formed star clusters out to the boundary of the soliton, thereby revealing the soliton radius directly with tidal streams\cite{Schive:2020,Dutta:2021}.

\begin{acknowledgments}
We thank Justin Schive for useful comments
and for providing the simulation profiles.\hfill \break

We are committed to transparency and collaboration in scientific research. Therefore, we extend an offer to provide our code and compiled dataset of published data to interested parties upon request to the authors. This gesture aims to facilitate openness and aid in furthering scientific inquiry. Please feel free to reach out to us for access to these resources.
\end{acknowledgments}

\appendix
\begin{center}
    
\textbf{\huge Appendix:}

\end{center}

\section{Data Analysis and Results}

We have explored the full range of relevant parameter space with the Monte Carlo Markov Chain (MCMC) technique based on the Metropolis-Hastings sampling  algorithm, to obtain the core radius that characterises fully the soliton profile, $r_c$, and the scale radius and the normalization of the stellar density profile,$ r_{s*} $ and $ \rho_{0*} $, and the also the transition radius $r_t$ between the core and halo profiles.The MCMC function was built in Mathematica software “Statistics MCMC` as the basis of our own MCMC function code. We allow for an adaptive step size in order to reach an acceptance rate between 20\% and 50\%, computing four chains of 10000 iterations for each variable($r_c$, $r_t$, $ \rho_{0*} $ and $ r_{s*} $) in each galaxy. Each chain's serial correlation was checked by correlograms (ACF plots), ensuring that the autocorrelation of the terms drop to zero before 250 lags \cite{Roy:2020}. We ensure the convergence relying on the Gelman-Rubin criteria adopting a Max Gelman–Rubin Rc below to 1.2 \cite{Roy:2020}. Once the convergence criteria are satisfied the chains are combined to compute the total likelihood, together with the 1D marginalized likelihood distribution with the corresponding the expectation value and  variance. The results are shown in Table~1 for the above free parameters and their uncertainties. We also show the covariances between the free parameters in Figures~6-10 for the isolated galaxies and for sextans, as the profiles of these galaxies may be regarded as unaffected by tidal interaction, as described in the text, reflecting intrinsic properties
in the context of $\psi$DM. We set flat priors for all the galaxies MCMC calculations, with the following uniform distribution for the two main parameters: $ r_c(kpc) \sim\mathcal{U}(0.1,\,0.75)$ and $ r_t(kpc) \sim\mathcal{U}(0.75,\,1.5)$ to span a wide range of $\psi$DM simulation expectations \cite{Schive:2014,Schive:2016}. We have adopted throughout a fixed boson mass of $\simeq 1.5\times 10^{-22}$ eV., close to previous estimates \cite{Schive:2014,Schive:2016}, which has no effect on the fitted core and halo scale lengths, but enters only in connection to the velocity dispersion predictions through the soliton to halo mass scaling relation, equation \ref{eq:sol_radius}, modestly affecting the shape of the velocity dispersion profile, as shown in figure 6.\\


\begin{table*}[h]
\centering
\begin{tabular}{|c|c|c|c|c|c|c|c|c|}

\hline

 Galaxy   & $r_c$  & $r_{t}$  & $r_{s*}$ &   Gap(Log) & Distance & Pericenter& $\sigma_{los}$& Ref \\
      & (kpc)&  (kpc) & (kpc)  & $\Delta_{C-H}$&(kpc)&(kpc)&(km/s)&-\\
\hline

\textcolor{blue}{Aquarius}  &$0.35^{+0.01}_{-0.01}$ & $1.25^{+0.07}_{-0.06}$  & $1.05^{+0.82}_{-0.64}$& 1.78&1071&-&-&-\\
\hline

\textcolor{blue}{Cetus}  &$0.36^{+0.02}_{-0.02}$ &$0.87^{+0.08}_{-0.07}$  & $0.24^{+0.14}_{-0.06}$& 1.34&775&-&-&-\\
\hline

\textcolor{blue}{Tucana}  &$0.25^{+0.01}_{-0.01}$ &$0.78^{+0.06}_{-0.06}$  & $1.05^{+0.90}_{-0.57}$ &1.5&887&-&$13.3^{+2.7}_{-2.3}$&\cite{Gregory:2019}\\

 \hline

\textcolor{blue}{Leo A}  &$0.43^{+0.02}_{-0.02}$  &$1.51^{+0.09}_{-0.08}$  & $0.73^{+0.49}_{-0.47}$& 1.76&800&-&-&-\\
\hline

\textcolor{red}{Sextans}   &$0.48^{+0.01}_{-0.01}$& $1.31^{+0.05}_{-0.06}$  & $1.61^{+0.51}_{-0.49}$&1.82&90&$71^{+11}_{-12}$&$7.0^{+1.3}_{-1.3}$&\cite{Cicuendez:2018}\\
\hline

\textcolor{red}{Phoenix} &$0.28^{+0.05}_{-0.06}$ & $1.27^{+0.01}_{-0.01}$  & $1.10^{+0.54}_{-0.55}$&3.44&415&$263^{+126}_{-219}$&$9.3^{+0.7}_{-0.7}$&\cite{Zaggia:2011}\\
\hline

\textcolor{red}{Leo I}  &$0.24^{+0.01}_{-0.01}$ & $1.30^{+0.08}_{-0.08}$  & $1.75^{+0.78}_{-0.96}$&3.43&250& $45^{+80}_{-34}$&$9.2^{+1.2}_{-1.2}$&\cite{Koch:2007}\\
\hline

\textcolor{red}{Draco}  &$0.17^{+0.01}_{-0.01}$ & $0.56^{+0.02}_{-0.02}$  & $0.1^{+0.09}_{-0.05}$ &2.32&80&$28^{+12}_{-7}$&$9.1^{+1.2}_{-1.2}$&\cite{Lokas:2005}\\
\hline
\textcolor{red}{Carina} &$0.21^{+0.01}_{-0.01}$ &  $0.81^{+0.04}_{-0.04}$  & $1.17^{+0.51}_{-0.61}$&2.26&101&$60^{+21}_{-16}$&-&-\\
\hline



\textcolor{red}{Sculptor}  &$0.20^{+0.01}_{-0.01}$  &$0.72^{+0.07}_{-0.07}$ &$0.12^{+0.25}_{-0.09}$&2.18&80&$51^{+15}_{-10}$&$9.2^{+1.4}_{-1.4}$&\cite{Walker:2009}\\

\hline

\textcolor{red}{Ursa Minor}   &$0.28^{+0.01}_{-0.01}$ & $0.96^{+0.05}_{-0.04}$  & $0.52^{+0.90}_{-0.40}$ &2.14&66&$29^{+8}_{-6}$&$9.5^{+1.2}_{-1.2}$&\cite{Walker:2009}\\
\hline

\textcolor{red}{Leo II}   &$0.14^{+0.01}_{-0.01}$ &  $0.60^{+0.03}_{-0.03}$  &$1.34^{+0.42}_{-0.55}$&2.45&210&$45^{+121}_{-30}$&$6.6^{+0.7}_{-0.7}$&\cite{Walker:2009}\\
\hline
 \hline
\end{tabular}
\caption{Observations and $\psi$DM profile fits. Column 1: Dwarf galaxy colour coded as in figure 2b, Column 2: Core radius  
$r_c$,  Column 3: Transition point $r_t$, Column 4: Stellar scale radius $r_{s*}$, Column 5: Gap $\Delta_{C-H}$, Column 6: Distances from Milky Way center \& Column 7: Pericenter determined from GAIA \cite{Fritz:2018},  Column 8: Mean dispersion velocity $\sigma_{los}$ from \cite{McConnachie:2020},  Column 9: References of the profiles of figure 6.}

\label{tabla:1}
\end{table*}

\begin{figure*}[p]
	\centering
	\includegraphics[width=2.00\columnwidth,height=15cm]{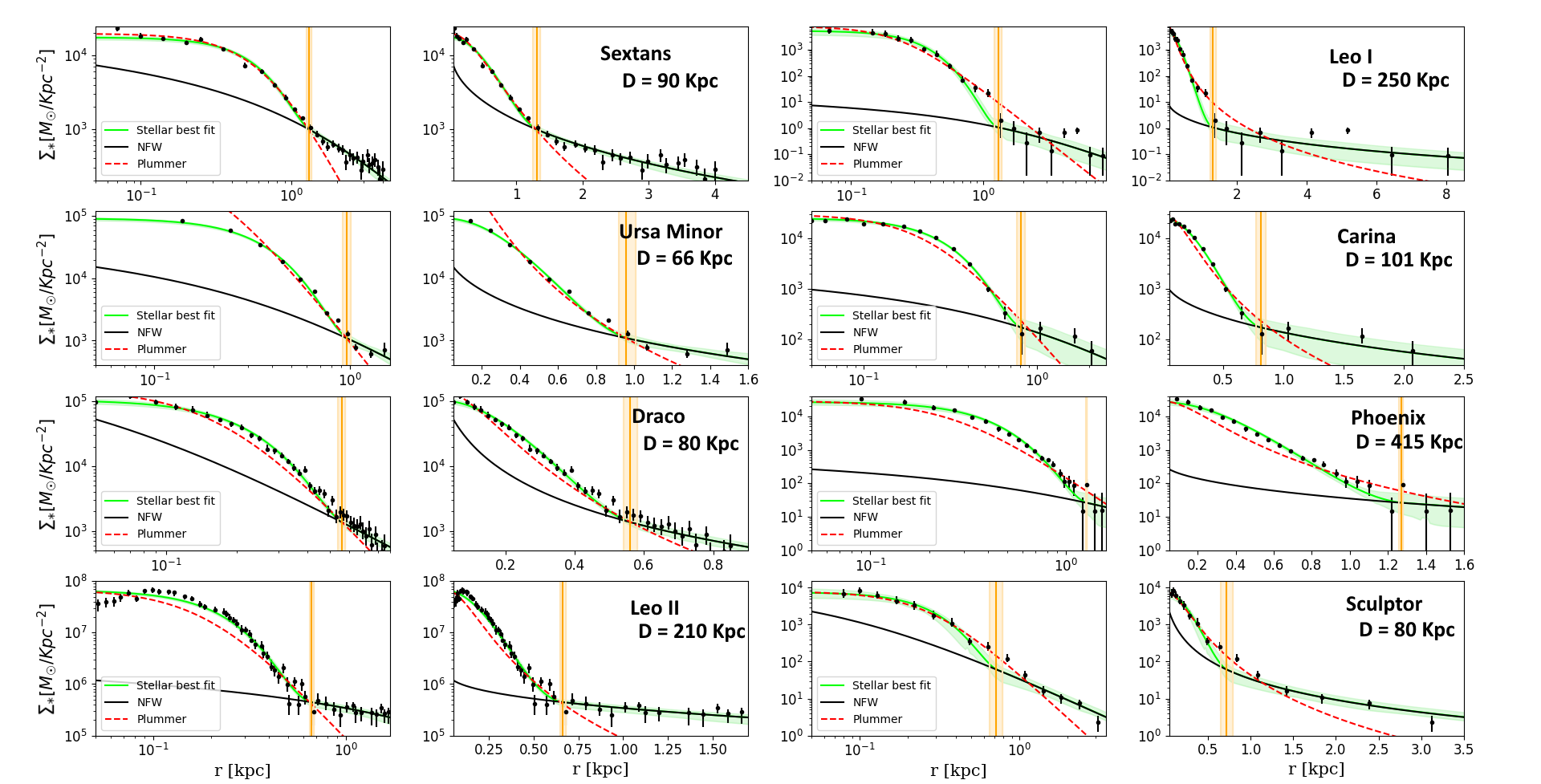}
	\caption{Stellar density profiles of orbiting dwarf galaxies listed on table 1. The green profile shows the 95\% uncertainty on the fitted profile obtained from the MCMC simulation. with the model transition radius $r_t$ and uncertainty marked 
	as the vertical orange bar, separating the core and halo regions. Notice the highly extended halos for Carina, Leo I, Sculptor, Sextans and Leo II were after subtracting the background level, stars can be detected to even 8kpc in the case of Leo I. The red dashed line represents the standard Plummer profile with the $r_{half}$ of each galaxy as the only free parameter, observe how  is not sufficiently pronounced for almost all the galaxies in contrast to the soliton. References for the data are: Carina\cite{Frinchaboy:2012},Ursa Minor\cite{Martinez-Delgado:2001}, Leo I\cite{Sohn:2007}, Leo II\cite{Coleman:2007}, Phoenix\cite{Battaglia:2012}, Sculptor \cite{Frinchaboy:2012}, Sextans \cite{Okamoto:2017} and Draco\cite{Wilkinson:2004}).}\label{fig5}
\end{figure*}



\begin{figure*}[htp]
	\centering
	\includegraphics[width=2\columnwidth,height=11.0cm]{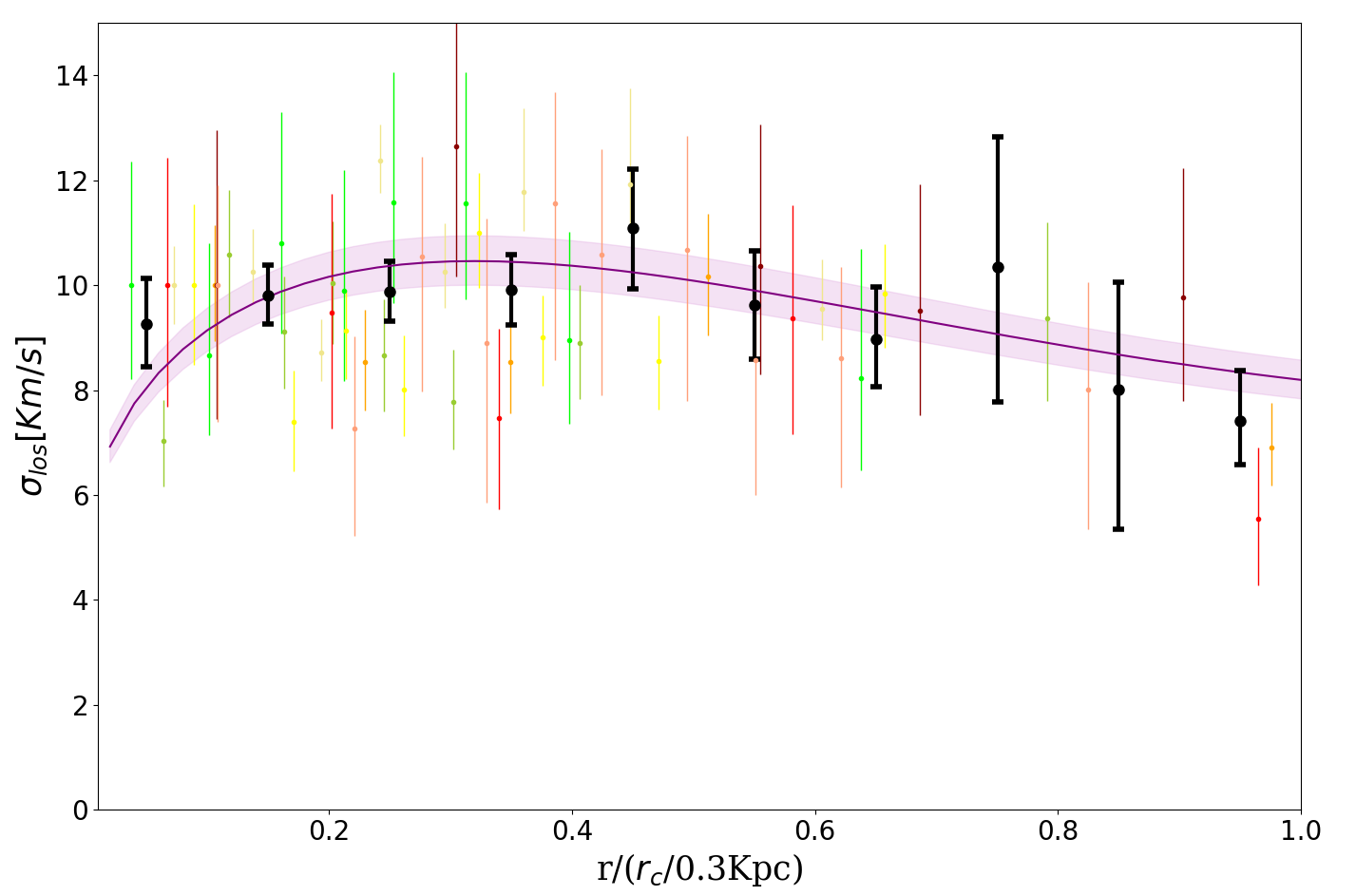}
	\caption{Comparison of the predicted mean velocity dispersion of the $\psi$DM profile (purple model curve), with the mean observed velocity dispersion shown as black data points with errors averaged over the eight dwarfs with well resolved dynamical data, in units of the core radius from fitting the star counts (listed in Table 1). The individual dispersion data are also shown for each dwarf, (coloured in the same way as Figures 2 \& 3) and normalised to the mean level for this comparison, demonstrating the dwarfs follow the general form expected for $\psi$DM, peaking  near the observed mean core radius of $\simeq 0.3$kpc, and then declining into the lower density halo. The mean level of the dispersion at the peak is $10km/s$ and together with the mean core radius of 0.3kpc may be used to obtain an approximate boson mass, via eqn A4 to $\simeq$ $1.5\times10^{-22}$eV.}\label{fig4}
\end{figure*}

\begin{figure*}[p]
	\centering
	\includegraphics[width=2.1\columnwidth,height=15cm]{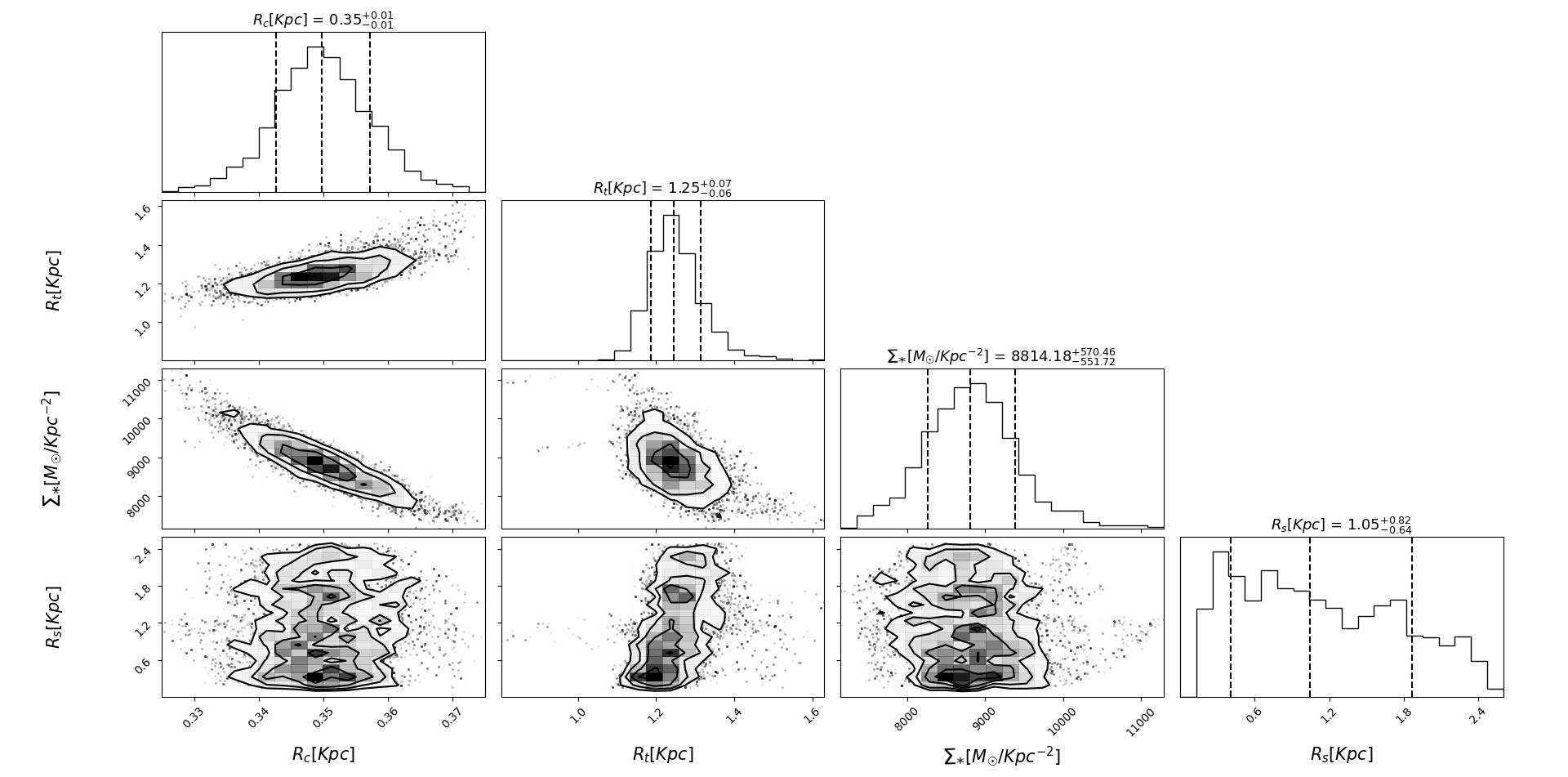}
	\caption{Aquarius: correlated distributions of free parameters from MCMC simulation. The core radius and transition radius is well defined here, despite the flat input priors, indicating a reliable result. The contours represent the 68\%, 95\%, and 99\% of confidence level. The best-fit parameter values are the medians, represented with the vertical red lines while the black ones show their errors. Picture created using the 'Corner.py' software.}\label{fig7}
\end{figure*}

\begin{figure*}[p]
	\centering
	\includegraphics[width=2.1\columnwidth,height=15cm]{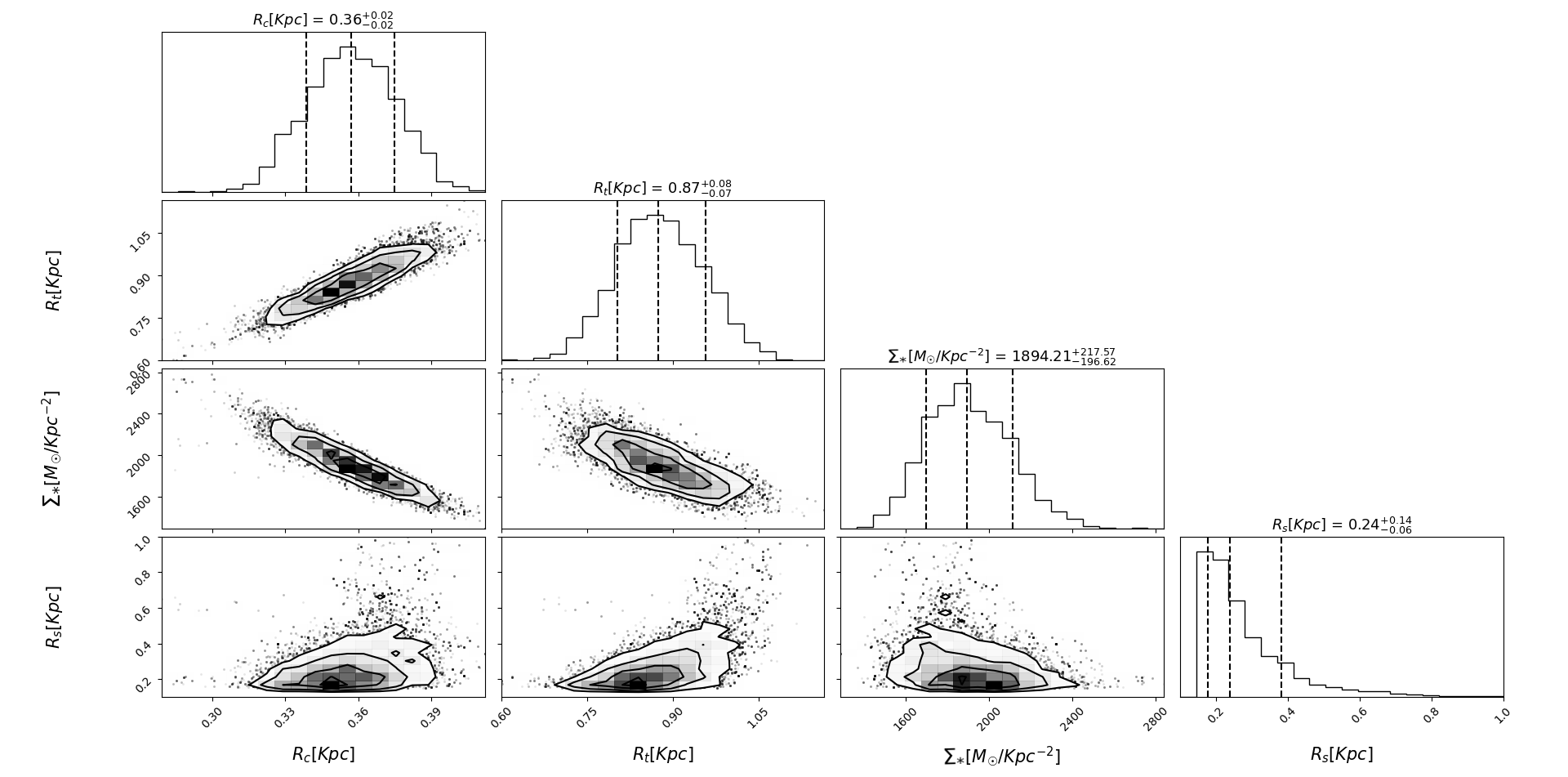}
	\caption{Cetus: correlated distributions of the free parameters.  As can be seen the core radius and transition radius are well defined despite the flat input priors, indicating a reliable result.The contours represent the 68\%, 95\%, and 99\% of confidence level. The best-fit parameter values are the medians, represented with the vertical red lines while the black ones show their errors. Picture created using the 'Corner.py' software.
	 }\label{fig9}
\end{figure*}

\begin{figure*}[p]
	\centering
	\includegraphics[width=2.1\columnwidth,height=15cm]{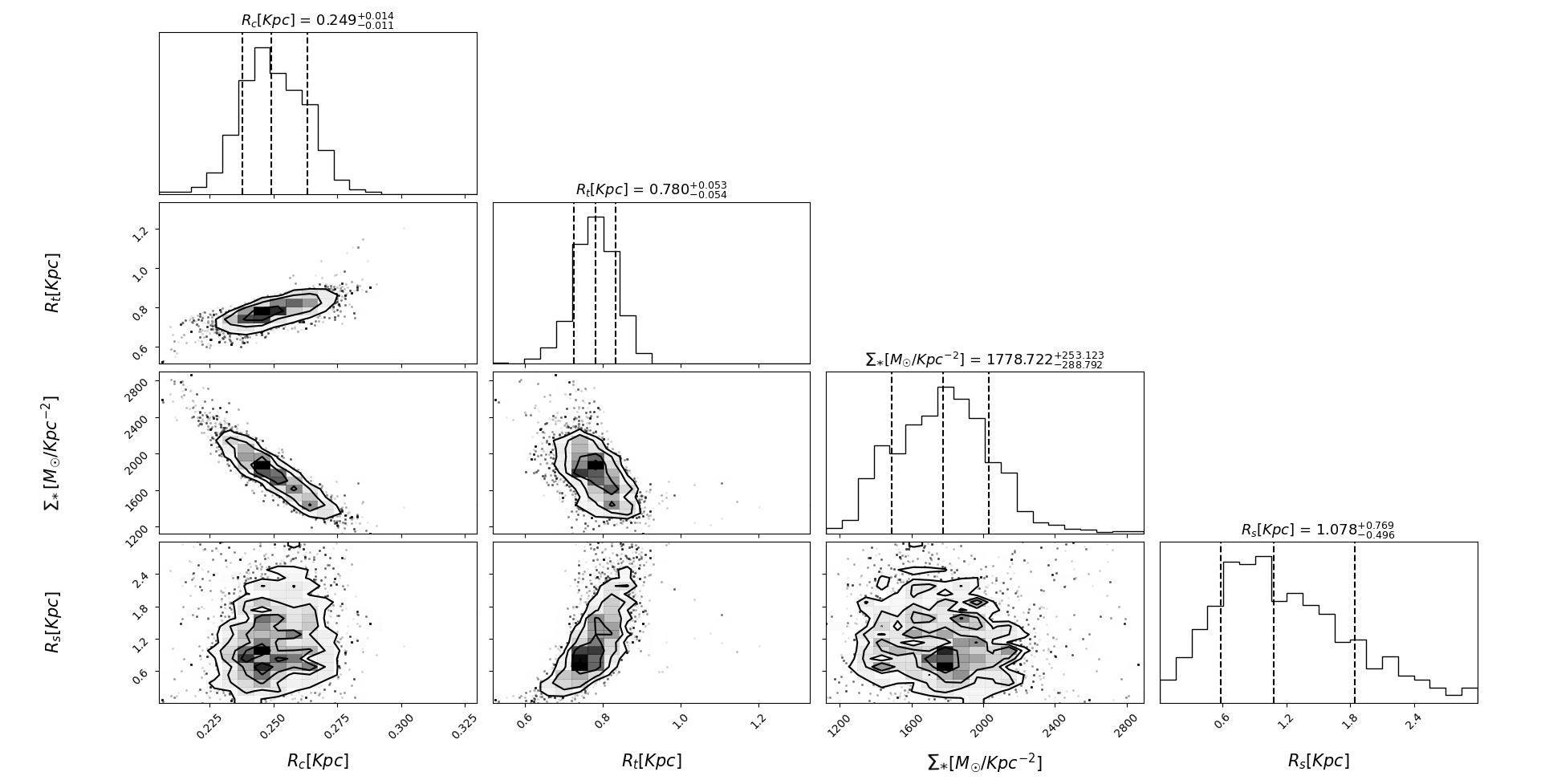}
	\caption{Tucana's correlated distributions of free parameters. 
	 As can be seen the core radius and transition radius are well defined despite the flat input priors, indicating a reliable result
	with a well constrained core and transition radius.The contours represent the 68\%, 95\%, and 99\% of confidence level. The best-fit parameter values are the medians, represented with the vertical red lines while the black ones show their errors. Picture created using the 'Corner.py' software.}\label{fig10}
\end{figure*}

\begin{figure*}[p]
	\centering
	\includegraphics[width=2.1\columnwidth,height=15cm]{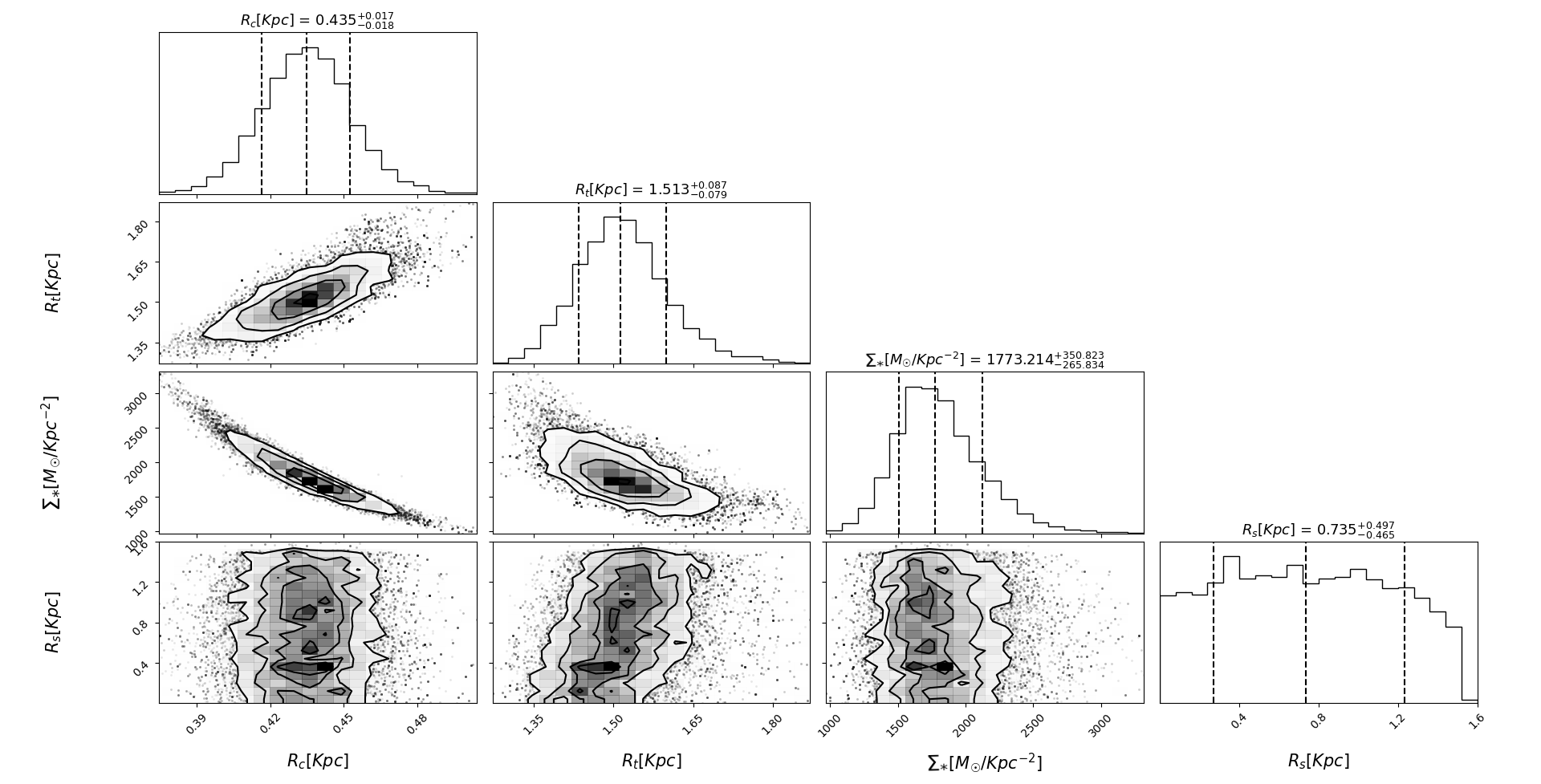}
	\caption{Leo A's correlated distributions of free parameters.
	 As can be seen the core radius and transition radius are well defined despite the flat input priors, indicating a reliable result.The contours represent the 68\%, 95\%, and 99\% of confidence level. The best-fit parameter values are the medians, represented with the vertical red lines while the black ones show their errors. Picture created using the 'Corner.py' software.
	 }\label{fig11}
\end{figure*}

\begin{figure*}[p]
	\centering
	\includegraphics[width=2.1\columnwidth,height=15cm]{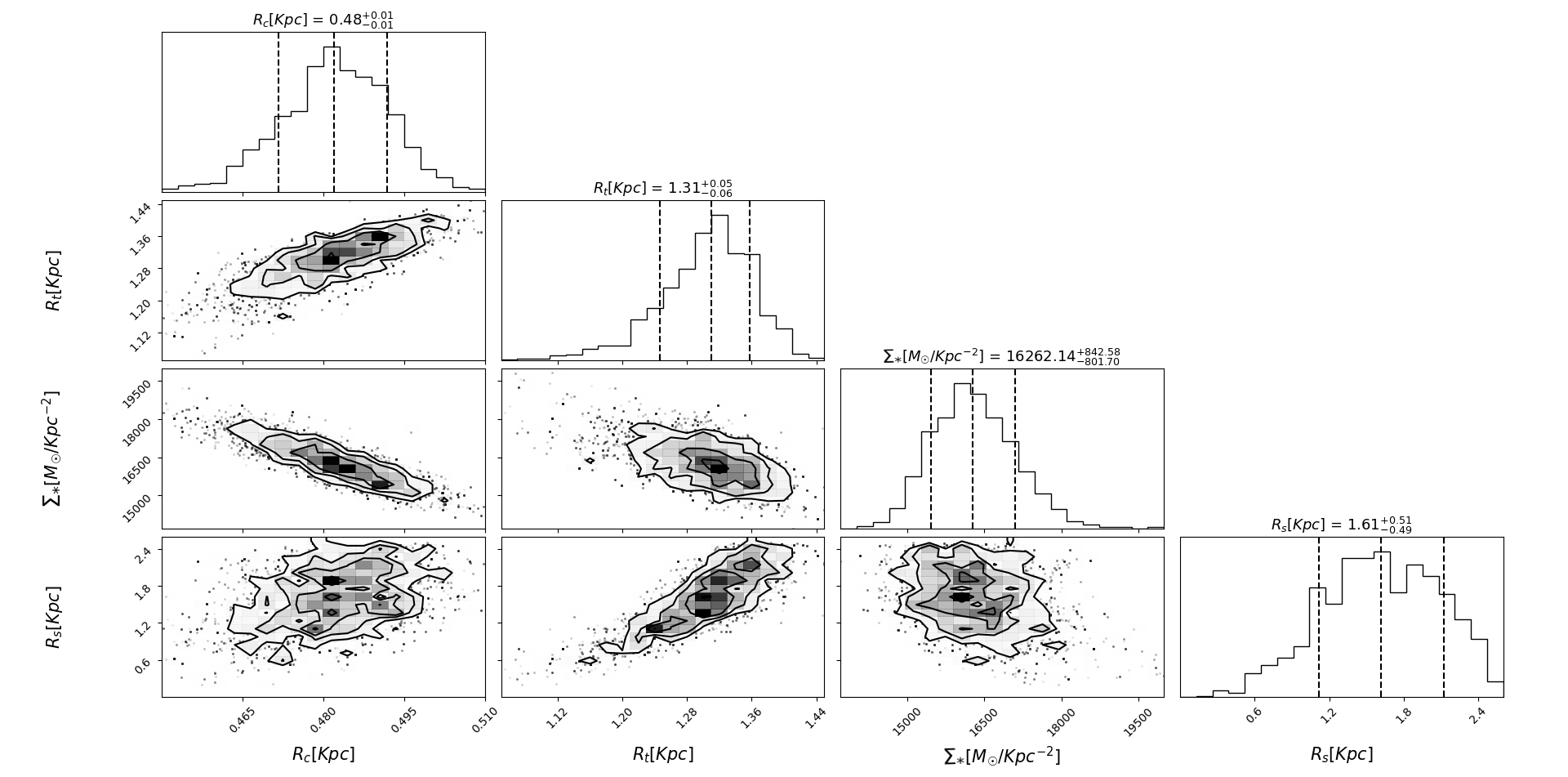}
	\caption{Sextans: correlated distributions of free parameters.
	 As can be seen the core radius and transition radius are well defined despite the flat input priors, indicating a reliable result.The contours represent the 68\%, 95\%, and 99\% of confidence level. The best-fit parameter values are the medians, represented with the vertical red lines while the black ones show their errors. Picture created using the 'Corner.py' software.
	 }\label{fig12}
\end{figure*}

\bibliography{Main}

\providecommand{\noopsort}[1]{}\providecommand{\singleletter}[1]{#1}%
\begin{thebibliography}{83}%
\makeatletter
\providecommand \@ifxundefined [1]{%
 \@ifx{#1\undefined}
}%
\providecommand \@ifnum [1]{%
 \ifnum #1\expandafter \@firstoftwo
 \else \expandafter \@secondoftwo
 \fi
}%
\providecommand \@ifx [1]{%
 \ifx #1\expandafter \@firstoftwo
 \else \expandafter \@secondoftwo
 \fi
}%
\providecommand \natexlab [1]{#1}%
\providecommand \enquote  [1]{``#1''}%
\providecommand \bibnamefont  [1]{#1}%
\providecommand \bibfnamefont [1]{#1}%
\providecommand \citenamefont [1]{#1}%
\providecommand \href@noop [0]{\@secondoftwo}%
\providecommand \href [0]{\begingroup \@sanitize@url \@href}%
\providecommand \@href[1]{\@@startlink{#1}\@@href}%
\providecommand \@@href[1]{\endgroup#1\@@endlink}%
\providecommand \@sanitize@url [0]{\catcode `\\12\catcode `\$12\catcode
  `\&12\catcode `\#12\catcode `\^12\catcode `\_12\catcode `\%12\relax}%
\providecommand \@@startlink[1]{}%
\providecommand \@@endlink[0]{}%
\providecommand \url  [0]{\begingroup\@sanitize@url \@url }%
\providecommand \@url [1]{\endgroup\@href {#1}{\urlprefix }}%
\providecommand \urlprefix  [0]{URL }%
\providecommand \Eprint [0]{\href }%
\providecommand \doibase [0]{https://doi.org/}%
\providecommand \selectlanguage [0]{\@gobble}%
\providecommand \bibinfo  [0]{\@secondoftwo}%
\providecommand \bibfield  [0]{\@secondoftwo}%
\providecommand \translation [1]{[#1]}%
\providecommand \BibitemOpen [0]{}%
\providecommand \bibitemStop [0]{}%
\providecommand \bibitemNoStop [0]{.\EOS\space}%
\providecommand \EOS [0]{\spacefactor3000\relax}%
\providecommand \BibitemShut  [1]{\csname bibitem#1\endcsname}%
\let\auto@bib@innerbib\@empty
\bibitem [{\citenamefont {Gregory}\ \emph {et~al.}(2019)\citenamefont
  {Gregory}, \citenamefont {Collins}, \citenamefont {Read}, \citenamefont
  {Irwin}, \citenamefont {Ibata}, \citenamefont {Martin}, \citenamefont
  {McConnachie},\ and\ \citenamefont {Weisz}}]{Gregory:2019}%
  \BibitemOpen
  \bibfield  {author} {\bibinfo {author} {\bibfnamefont {A.~L.}\ \bibnamefont
  {Gregory}}, \bibinfo {author} {\bibfnamefont {M.~L.~M.}\ \bibnamefont
  {Collins}}, \bibinfo {author} {\bibfnamefont {J.~I.}\ \bibnamefont {Read}},
  \bibinfo {author} {\bibfnamefont {M.~J.}\ \bibnamefont {Irwin}}, \bibinfo
  {author} {\bibfnamefont {R.~A.}\ \bibnamefont {Ibata}}, \bibinfo {author}
  {\bibfnamefont {N.~F.}\ \bibnamefont {Martin}}, \bibinfo {author}
  {\bibfnamefont {A.~W.}\ \bibnamefont {McConnachie}},\ and\ \bibinfo {author}
  {\bibfnamefont {D.~R.}\ \bibnamefont {Weisz}},\ }\bibfield  {title} {\bibinfo
  {title} {Kinematics of the tucana dwarf galaxy: an unusually dense dwarf in
  the local group.},\ }\href
  {https://doi.org/https://doi.org/10.1093/mnras/stz518} {\bibfield  {journal}
  {\bibinfo  {journal} {Mon. Not. R. Astron. Soc.}\ }\textbf {\bibinfo {volume}
  {485}},\ \bibinfo {pages} {2010} (\bibinfo {year} {2019})}\BibitemShut
  {NoStop}%
\bibitem [{\citenamefont {McConnachie}\ and\ \citenamefont
  {Irwin}(2006)}]{McConnachie:2006}%
  \BibitemOpen
  \bibfield  {author} {\bibinfo {author} {\bibfnamefont {A.~W.}\ \bibnamefont
  {McConnachie}}\ and\ \bibinfo {author} {\bibfnamefont {M.~J.}\ \bibnamefont
  {Irwin}},\ }\bibfield  {title} {\bibinfo {title} {Structural properties of
  the m31 dwarf spheroidal galaxies.},\ }\href
  {https://doi.org/https://doi.org/10.1111/j.1365-2966.2005.09806.x} {\bibfield
   {journal} {\bibinfo  {journal} {Mon. Not. R. Astron. Soc.}\ }\textbf
  {\bibinfo {volume} {365}},\ \bibinfo {pages} {1263} (\bibinfo {year}
  {2006})}\BibitemShut {NoStop}%
\bibitem [{\citenamefont {Kang}\ and\ \citenamefont
  {Ricotti}(2019)}]{Kang:2019}%
  \BibitemOpen
  \bibfield  {author} {\bibinfo {author} {\bibfnamefont {H.~D.}\ \bibnamefont
  {Kang}}\ and\ \bibinfo {author} {\bibfnamefont {M.}~\bibnamefont {Ricotti}},\
  }\bibfield  {title} {\bibinfo {title} {Ghostly haloes in dwarf galaxies:
  constraints on the star formation efficiency before reionization.},\ }\href
  {https://doi.org/https://doi.org/10.1093/mnras/stz1886} {\bibfield  {journal}
  {\bibinfo  {journal} {Mon. Not. R. Astron. Soc.}\ }\textbf {\bibinfo {volume}
  {488}},\ \bibinfo {pages} {2673} (\bibinfo {year} {2019})}\BibitemShut
  {NoStop}%
\bibitem [{\citenamefont {McConnachie}\ \emph {et~al.}(2006)\citenamefont
  {McConnachie}, \citenamefont {Arimoto}, \citenamefont {Irwin},\ and\
  \citenamefont {Tolstoy}}]{McConnachie:20062}%
  \BibitemOpen
  \bibfield  {author} {\bibinfo {author} {\bibfnamefont {A.~W.}\ \bibnamefont
  {McConnachie}}, \bibinfo {author} {\bibfnamefont {N.}~\bibnamefont
  {Arimoto}}, \bibinfo {author} {\bibfnamefont {M.}~\bibnamefont {Irwin}},\
  and\ \bibinfo {author} {\bibfnamefont {E.}~\bibnamefont {Tolstoy}},\
  }\bibfield  {title} {\bibinfo {title} {The stellar content of the isolated
  transition dwarf galaxy ddo210.},\ }\href
  {https://doi.org/https://doi.org/10.1111/j.1365-2966.2006.11053.x} {\bibfield
   {journal} {\bibinfo  {journal} {Mon. Not. R. Astron. Soc.}\ }\textbf
  {\bibinfo {volume} {373}},\ \bibinfo {pages} {715} (\bibinfo {year}
  {2006})}\BibitemShut {NoStop}%
\bibitem [{\citenamefont {Cyburt}\ \emph {et~al.}(2016)\citenamefont {Cyburt},
  \citenamefont {Fields}, \citenamefont {Olive},\ and\ \citenamefont
  {Yeh}}]{Cyburt:2016}%
  \BibitemOpen
  \bibfield  {author} {\bibinfo {author} {\bibfnamefont {R.~H.}\ \bibnamefont
  {Cyburt}}, \bibinfo {author} {\bibfnamefont {B.~D.}\ \bibnamefont {Fields}},
  \bibinfo {author} {\bibfnamefont {K.~A.}\ \bibnamefont {Olive}},\ and\
  \bibinfo {author} {\bibfnamefont {T.-H.}\ \bibnamefont {Yeh}},\ }\bibfield
  {title} {\bibinfo {title} {Big bang nucleosynthesis: Present status.},\
  }\bibfield  {journal} {\bibinfo  {journal} {Rev.of Modern Phys.}\ }\textbf
  {\bibinfo {volume} {88}},\ \href
  {https://doi.org/https://doi.org/10.1103/RevModPhys.88.015004}
  {https://doi.org/10.1103/RevModPhys.88.015004} (\bibinfo {year}
  {2016})\BibitemShut {NoStop}%
\bibitem [{\citenamefont {Collaboration}(2016)}]{Planck:2016}%
  \BibitemOpen
  \bibfield  {author} {\bibinfo {author} {\bibfnamefont {P.}~\bibnamefont
  {Collaboration}},\ }\bibfield  {title} {\bibinfo {title} {Planck 2015
  results. xiii. cosmological parameters.},\ }\href
  {https://doi.org/https://doi.org/10.1051/0004-6361/201525830} {\bibfield
  {journal} {\bibinfo  {journal} {Astro. $\&$ Astrophys.}\ }\textbf {\bibinfo
  {volume} {594}},\ \bibinfo {pages} {63} (\bibinfo {year} {2016})}\BibitemShut
  {NoStop}%
\bibitem [{\citenamefont {Markevitch}\ \emph {et~al.}(2004)\citenamefont
  {Markevitch}, \citenamefont {Gonzalez}, \citenamefont {Clowe}, \citenamefont
  {Vikhlinin}, \citenamefont {Forman}, \citenamefont {Jones}, \citenamefont
  {Murray},\ and\ \citenamefont {Tucker}}]{Markevitch:2004}%
  \BibitemOpen
  \bibfield  {author} {\bibinfo {author} {\bibfnamefont {M.}~\bibnamefont
  {Markevitch}}, \bibinfo {author} {\bibfnamefont {A.~H.}\ \bibnamefont
  {Gonzalez}}, \bibinfo {author} {\bibfnamefont {D.}~\bibnamefont {Clowe}},
  \bibinfo {author} {\bibfnamefont {A.}~\bibnamefont {Vikhlinin}}, \bibinfo
  {author} {\bibfnamefont {W.}~\bibnamefont {Forman}}, \bibinfo {author}
  {\bibfnamefont {C.}~\bibnamefont {Jones}}, \bibinfo {author} {\bibfnamefont
  {S.}~\bibnamefont {Murray}},\ and\ \bibinfo {author} {\bibfnamefont
  {W.}~\bibnamefont {Tucker}},\ }\bibfield  {title} {\bibinfo {title} {Direct
  constraints on the dark matter self-interaction cross section from the
  merging galaxy cluster 1e 0657-56.},\ }\href
  {https://doi.org/https://doi.org/10.1086/383178} {\bibfield  {journal}
  {\bibinfo  {journal} {Astrophys. J.}\ }\textbf {\bibinfo {volume} {606}},\
  \bibinfo {pages} {819} (\bibinfo {year} {2004})}\BibitemShut {NoStop}%
\bibitem [{\citenamefont {Clowe}\ \emph {et~al.}(2006)\citenamefont {Clowe},
  \citenamefont {Bradač}, \citenamefont {Gonzalez}, \citenamefont
  {Markevitch}, \citenamefont {Randall}, \citenamefont {Jones},\ and\
  \citenamefont {Zaritsky}}]{Clowe:2006}%
  \BibitemOpen
  \bibfield  {author} {\bibinfo {author} {\bibfnamefont {D.}~\bibnamefont
  {Clowe}}, \bibinfo {author} {\bibfnamefont {M.}~\bibnamefont {Bradač}},
  \bibinfo {author} {\bibfnamefont {A.~H.}\ \bibnamefont {Gonzalez}}, \bibinfo
  {author} {\bibfnamefont {M.}~\bibnamefont {Markevitch}}, \bibinfo {author}
  {\bibfnamefont {S.~W.}\ \bibnamefont {Randall}}, \bibinfo {author}
  {\bibfnamefont {C.}~\bibnamefont {Jones}},\ and\ \bibinfo {author}
  {\bibfnamefont {D.}~\bibnamefont {Zaritsky}},\ }\bibfield  {title} {\bibinfo
  {title} {A direct empirical proof of the existence of dark matter.},\ }\href
  {https://doi.org/https://doi.org/10.1086/508162} {\bibfield  {journal}
  {\bibinfo  {journal} {Astrophys. J.}\ }\textbf {\bibinfo {volume} {648}},\
  \bibinfo {pages} {L109} (\bibinfo {year} {2006})}\BibitemShut {NoStop}%
\bibitem [{\citenamefont {Aprile}(2018)}]{Aprile:2018}%
  \BibitemOpen
  \bibfield  {author} {\bibinfo {author} {\bibfnamefont {E.~a. X.~C.}\
  \bibnamefont {Aprile}},\ }\bibfield  {title} {\bibinfo {title} {Dark matter
  search results from a one ton-year exposure of xenon1t.},\ }\bibfield
  {journal} {\bibinfo  {journal} {Phys. Rev. Lett.}\ }\textbf {\bibinfo
  {volume} {121}},\ \href
  {https://doi.org/https://doi.org/10.1103/PhysRevLett.121.111302}
  {https://doi.org/10.1103/PhysRevLett.121.111302} (\bibinfo {year}
  {2018})\BibitemShut {NoStop}%
\bibitem [{\citenamefont {Moore}(1994)}]{Moore:1994}%
  \BibitemOpen
  \bibfield  {author} {\bibinfo {author} {\bibfnamefont {B.}~\bibnamefont
  {Moore}},\ }\bibfield  {title} {\bibinfo {title} {Evidence against
  dissipation-less dark matter from observations of galaxy haloes.},\ }\href
  {https://doi.org/https://doi.org/10.1038/370629a0} {\bibfield  {journal}
  {\bibinfo  {journal} {Nature}\ }\textbf {\bibinfo {volume} {370}},\ \bibinfo
  {pages} {629} (\bibinfo {year} {1994})}\BibitemShut {NoStop}%
\bibitem [{\citenamefont {de~Blok}(2010)}]{deBlok:2010}%
  \BibitemOpen
  \bibfield  {author} {\bibinfo {author} {\bibfnamefont {W.~J.~G.}\
  \bibnamefont {de~Blok}},\ }\bibfield  {title} {\bibinfo {title} {The
  core-cusp problem.},\ }\bibfield  {journal} {\bibinfo  {journal} {Adv. in
  Astro.}\ }\href {https://doi.org/https://doi.org/10.1155/2010/789293}
  {https://doi.org/10.1155/2010/789293} (\bibinfo {year} {2010})\BibitemShut
  {NoStop}%
\bibitem [{\citenamefont {Marsh}\ and\ \citenamefont
  {Silk}(2014)}]{Marsh:2014}%
  \BibitemOpen
  \bibfield  {author} {\bibinfo {author} {\bibfnamefont {D.~J.~E.}\
  \bibnamefont {Marsh}}\ and\ \bibinfo {author} {\bibfnamefont
  {J.}~\bibnamefont {Silk}},\ }\bibfield  {title} {\bibinfo {title} {A model
  for halo formation with axion mixed dark matter.},\ }\href
  {https://doi.org/https://doi.org/10.1093/mnras/stt2079} {\bibfield  {journal}
  {\bibinfo  {journal} {Mon. Not. R. Astron. Soc.}\ }\textbf {\bibinfo {volume}
  {437}},\ \bibinfo {pages} {2652} (\bibinfo {year} {2014})}\BibitemShut
  {NoStop}%
\bibitem [{\citenamefont {Klypin}\ \emph {et~al.}(1999)\citenamefont {Klypin},
  \citenamefont {Kravtsov},\ and\ \citenamefont {Valenzuela}}]{Klypin:1999}%
  \BibitemOpen
  \bibfield  {author} {\bibinfo {author} {\bibfnamefont {A.}~\bibnamefont
  {Klypin}}, \bibinfo {author} {\bibfnamefont {A.~V.}\ \bibnamefont
  {Kravtsov}},\ and\ \bibinfo {author} {\bibfnamefont {F.}~\bibnamefont
  {Valenzuela}, \bibfnamefont {O.~Prada}},\ }\bibfield  {title} {\bibinfo
  {title} {Where are the missing galactic satellites?},\ }\href
  {https://doi.org/https://doi.org/10.1086/307643} {\bibfield  {journal}
  {\bibinfo  {journal} {Astrophys. J.}\ }\textbf {\bibinfo {volume} {522}},\
  \bibinfo {pages} {82} (\bibinfo {year} {1999})}\BibitemShut {NoStop}%
\bibitem [{\citenamefont {Safarzadeh}\ and\ \citenamefont
  {Loeb}(2021)}]{Safarzadeh:2021}%
  \BibitemOpen
  \bibfield  {author} {\bibinfo {author} {\bibfnamefont {M.}~\bibnamefont
  {Safarzadeh}}\ and\ \bibinfo {author} {\bibfnamefont {A.}~\bibnamefont
  {Loeb}},\ }\bibfield  {title} {\bibinfo {title} {A new challenge for dark
  matter models.},\ }\bibfield  {journal} {\bibinfo  {journal} {Arxiv}\ }\href
  {https://doi.org/https://arxiv.org/pdf/2107.03478.pdf}
  {https://arxiv.org/pdf/2107.03478.pdf} (\bibinfo {year} {2021})\BibitemShut
  {NoStop}%
\bibitem [{\citenamefont {Schive}\ \emph
  {et~al.}(2014{\natexlab{a}})\citenamefont {Schive}, \citenamefont {Chiueh},\
  and\ \citenamefont {Broadhurst}}]{Schive:2014}%
  \BibitemOpen
  \bibfield  {author} {\bibinfo {author} {\bibfnamefont {H.-Y.}\ \bibnamefont
  {Schive}}, \bibinfo {author} {\bibfnamefont {T.}~\bibnamefont {Chiueh}},\
  and\ \bibinfo {author} {\bibfnamefont {T.}~\bibnamefont {Broadhurst}},\
  }\bibfield  {title} {\bibinfo {title} {Cosmic structure as the quantum
  interference of a coherent dark wave.},\ }\href
  {https://doi.org/https://doi.org/10.1038/nphys2996} {\bibfield  {journal}
  {\bibinfo  {journal} {Nature Phys}\ }\textbf {\bibinfo {volume} {10}},\
  \bibinfo {pages} {496} (\bibinfo {year} {2014}{\natexlab{a}})}\BibitemShut
  {NoStop}%
\bibitem [{\citenamefont {Schive}\ \emph
  {et~al.}(2014{\natexlab{b}})\citenamefont {Schive}, \citenamefont {Liao},
  \citenamefont {Woo}, \citenamefont {Wong}, \citenamefont {Chiueh},
  \citenamefont {Broadhurst},\ and\ \citenamefont {Hwang}}]{Schive:20142}%
  \BibitemOpen
  \bibfield  {author} {\bibinfo {author} {\bibfnamefont {H.-Y.}\ \bibnamefont
  {Schive}}, \bibinfo {author} {\bibfnamefont {M.-H.}\ \bibnamefont {Liao}},
  \bibinfo {author} {\bibfnamefont {T.-P.}\ \bibnamefont {Woo}}, \bibinfo
  {author} {\bibfnamefont {S.-K.}\ \bibnamefont {Wong}}, \bibinfo {author}
  {\bibfnamefont {T.}~\bibnamefont {Chiueh}}, \bibinfo {author} {\bibfnamefont
  {T.}~\bibnamefont {Broadhurst}},\ and\ \bibinfo {author} {\bibfnamefont
  {W.~Y.~P.}\ \bibnamefont {Hwang}},\ }\bibfield  {title} {\bibinfo {title}
  {Understanding the core-halo relation of quantum wave dark matter from 3d
  simulations.},\ }\bibfield  {journal} {\bibinfo  {journal} {Phys. Rev.
  Lett.}\ }\textbf {\bibinfo {volume} {113}},\ \href
  {https://doi.org/https://doi.org/10.1103/PhysRevLett.113.261302}
  {https://doi.org/10.1103/PhysRevLett.113.261302} (\bibinfo {year}
  {2014}{\natexlab{b}})\BibitemShut {NoStop}%
\bibitem [{\citenamefont {Schive}\ \emph {et~al.}(2016)\citenamefont {Schive},
  \citenamefont {Chiueh}, \citenamefont {Broadhurst},\ and\ \citenamefont
  {Huang}}]{Schive:2016}%
  \BibitemOpen
  \bibfield  {author} {\bibinfo {author} {\bibfnamefont {H.-Y.}\ \bibnamefont
  {Schive}}, \bibinfo {author} {\bibfnamefont {T.}~\bibnamefont {Chiueh}},
  \bibinfo {author} {\bibfnamefont {T.}~\bibnamefont {Broadhurst}},\ and\
  \bibinfo {author} {\bibfnamefont {K.-W.}\ \bibnamefont {Huang}},\ }\bibfield
  {title} {\bibinfo {title} {Contrasting galaxy formation from quantum wave
  dark matter.},\ }\href
  {https://doi.org/https://doi.org/10.3847/0004-637X/818/1/89} {\bibfield
  {journal} {\bibinfo  {journal} {Astrophys. J.}\ }\textbf {\bibinfo {volume}
  {818}},\ \bibinfo {pages} {14} (\bibinfo {year} {2016})}\BibitemShut
  {NoStop}%
\bibitem [{\citenamefont {Chiti}\ \emph {et~al.}(2021)\citenamefont {Chiti},
  \citenamefont {Frebel}, \citenamefont {Simon}, \citenamefont {Erkal},
  \citenamefont {Chang}, \citenamefont {Necib}, \citenamefont {Ji},
  \citenamefont {Jerjen}, \citenamefont {Kim},\ and\ \citenamefont
  {Norris}}]{Chiti:2021}%
  \BibitemOpen
  \bibfield  {author} {\bibinfo {author} {\bibfnamefont {A.}~\bibnamefont
  {Chiti}}, \bibinfo {author} {\bibfnamefont {A.}~\bibnamefont {Frebel}},
  \bibinfo {author} {\bibfnamefont {J.~D.}\ \bibnamefont {Simon}}, \bibinfo
  {author} {\bibfnamefont {D.}~\bibnamefont {Erkal}}, \bibinfo {author}
  {\bibfnamefont {L.~J.}\ \bibnamefont {Chang}}, \bibinfo {author}
  {\bibfnamefont {L.}~\bibnamefont {Necib}}, \bibinfo {author} {\bibfnamefont
  {A.~P.}\ \bibnamefont {Ji}}, \bibinfo {author} {\bibfnamefont
  {H.}~\bibnamefont {Jerjen}}, \bibinfo {author} {\bibfnamefont
  {D.}~\bibnamefont {Kim}},\ and\ \bibinfo {author} {\bibfnamefont {J.~E.}\
  \bibnamefont {Norris}},\ }\bibfield  {title} {\bibinfo {title} {An extended
  halo around an ancient dwarf galaxy.},\ }\bibfield  {journal} {\bibinfo
  {journal} {Nat. Astro.}\ }\href
  {https://doi.org/https://doi.org/10.1038/s41550-020-01285-w}
  {https://doi.org/10.1038/s41550-020-01285-w} (\bibinfo {year}
  {2021})\BibitemShut {NoStop}%
\bibitem [{\citenamefont {Collins}\ \emph {et~al.}(2021)\citenamefont
  {Collins}, \citenamefont {Read}, \citenamefont {Ibata}, \citenamefont {Rich},
  \citenamefont {Martin}, \citenamefont {Peñarrubia}, \citenamefont {Chapman},
  \citenamefont {Tollerud},\ and\ \citenamefont {Weisz}}]{Collins:2021}%
  \BibitemOpen
  \bibfield  {author} {\bibinfo {author} {\bibfnamefont {M.~L.~M.}\
  \bibnamefont {Collins}}, \bibinfo {author} {\bibfnamefont {J.~I.}\
  \bibnamefont {Read}}, \bibinfo {author} {\bibfnamefont {R.~A.}\ \bibnamefont
  {Ibata}}, \bibinfo {author} {\bibfnamefont {R.~M.}\ \bibnamefont {Rich}},
  \bibinfo {author} {\bibfnamefont {N.~F.}\ \bibnamefont {Martin}}, \bibinfo
  {author} {\bibfnamefont {J.}~\bibnamefont {Peñarrubia}}, \bibinfo {author}
  {\bibfnamefont {S.~C.}\ \bibnamefont {Chapman}}, \bibinfo {author}
  {\bibfnamefont {E.~J.}\ \bibnamefont {Tollerud}},\ and\ \bibinfo {author}
  {\bibfnamefont {D.~R.}\ \bibnamefont {Weisz}},\ }\bibfield  {title} {\bibinfo
  {title} {Andromeda xxi -- a dwarf galaxy in a low density dark matter halo},\
  }\bibfield  {journal} {\bibinfo  {journal} {ArXiv}\ }\href
  {https://doi.org/https://arxiv.org/pdf/2102.11890.pdf}
  {https://arxiv.org/pdf/2102.11890.pdf} (\bibinfo {year} {2021})\BibitemShut
  {NoStop}%
\bibitem [{\citenamefont {Torrealba}\ \emph {et~al.}(2016)\citenamefont
  {Torrealba}, \citenamefont {Koposov}, \citenamefont {Belokurov},\ and\
  \citenamefont {Irwin}}]{Torrealba:2016}%
  \BibitemOpen
  \bibfield  {author} {\bibinfo {author} {\bibfnamefont {G.}~\bibnamefont
  {Torrealba}}, \bibinfo {author} {\bibfnamefont {S.~E.}\ \bibnamefont
  {Koposov}}, \bibinfo {author} {\bibfnamefont {V.}~\bibnamefont {Belokurov}},\
  and\ \bibinfo {author} {\bibfnamefont {M.}~\bibnamefont {Irwin}},\ }\bibfield
   {title} {\bibinfo {title} {The feeble giant. discovery of a large and
  diffuse milky way dwarf galaxy in the constellation of crater},\ }\href
  {https://doi.org/https://doi.org/10.1093/mnras/stw733} {\bibfield  {journal}
  {\bibinfo  {journal} {Mon. Not. R. Astron. Soc.}\ }\textbf {\bibinfo {volume}
  {459}},\ \bibinfo {pages} {2370} (\bibinfo {year} {2016})}\BibitemShut
  {NoStop}%
\bibitem [{\citenamefont {Torrealba}\ \emph {et~al.}(2019)\citenamefont
  {Torrealba}, \citenamefont {Belokurov}, \citenamefont {Koposov},
  \citenamefont {Li}, \citenamefont {Walker}, \citenamefont {Sanders},
  \citenamefont {Geringer-Sameth}, \citenamefont {Zucker}, \citenamefont
  {Kuehn}, \citenamefont {Evans},\ and\ \citenamefont
  {Dehnen}}]{Torrealba:2019}%
  \BibitemOpen
  \bibfield  {author} {\bibinfo {author} {\bibfnamefont {G.}~\bibnamefont
  {Torrealba}}, \bibinfo {author} {\bibfnamefont {V.}~\bibnamefont
  {Belokurov}}, \bibinfo {author} {\bibfnamefont {S.~E.}\ \bibnamefont
  {Koposov}}, \bibinfo {author} {\bibfnamefont {T.~S.}\ \bibnamefont {Li}},
  \bibinfo {author} {\bibfnamefont {M.~G.}\ \bibnamefont {Walker}}, \bibinfo
  {author} {\bibfnamefont {J.~L.}\ \bibnamefont {Sanders}}, \bibinfo {author}
  {\bibfnamefont {A.}~\bibnamefont {Geringer-Sameth}}, \bibinfo {author}
  {\bibfnamefont {D.~B.}\ \bibnamefont {Zucker}}, \bibinfo {author}
  {\bibfnamefont {K.}~\bibnamefont {Kuehn}}, \bibinfo {author} {\bibfnamefont
  {N.~W.}\ \bibnamefont {Evans}},\ and\ \bibinfo {author} {\bibfnamefont
  {W.}~\bibnamefont {Dehnen}},\ }\bibfield  {title} {\bibinfo {title} {The
  hidden giant: discovery of an enormous galactic dwarf satellite in gaia
  dr2.},\ }\href {https://doi.org/https://doi.org/10.1093/mnras/stz1624}
  {\bibfield  {journal} {\bibinfo  {journal} {Mon. Not. R. Astron. Soc.}\
  }\textbf {\bibinfo {volume} {488}},\ \bibinfo {pages} {2743} (\bibinfo {year}
  {2019})}\BibitemShut {NoStop}%
\bibitem [{\citenamefont {Wilkinson}\ \emph {et~al.}(2004)\citenamefont
  {Wilkinson}, \citenamefont {Kleyna}, \citenamefont {Evans}, \citenamefont
  {Gilmore}, \citenamefont {Irwin},\ and\ \citenamefont
  {Grebel}}]{Wilkinson:2004}%
  \BibitemOpen
  \bibfield  {author} {\bibinfo {author} {\bibfnamefont {M.~I.}\ \bibnamefont
  {Wilkinson}}, \bibinfo {author} {\bibfnamefont {J.~T.}\ \bibnamefont
  {Kleyna}}, \bibinfo {author} {\bibfnamefont {N.~W.}\ \bibnamefont {Evans}},
  \bibinfo {author} {\bibfnamefont {G.~F.}\ \bibnamefont {Gilmore}}, \bibinfo
  {author} {\bibfnamefont {M.~J.}\ \bibnamefont {Irwin}},\ and\ \bibinfo
  {author} {\bibfnamefont {E.~K.}\ \bibnamefont {Grebel}},\ }\bibfield  {title}
  {\bibinfo {title} {Kinematically cold populations at large radii in the draco
  and ursa minor dwarf spheroidal galaxies.},\ }\href
  {https://doi.org/https://doi.org/10.1086/423619} {\bibfield  {journal}
  {\bibinfo  {journal} {Astrophys. J.}\ }\textbf {\bibinfo {volume} {611}},\
  \bibinfo {pages} {L21} (\bibinfo {year} {2004})}\BibitemShut {NoStop}%
\bibitem [{\citenamefont {Fabrizio}\ \emph {et~al.}(2016)\citenamefont
  {Fabrizio}, \citenamefont {Bono}, \citenamefont {Nonino}, \citenamefont
  {Łokas}, \citenamefont {Ferraro}, \citenamefont {Iannicola}, \citenamefont
  {Buonanno}, \citenamefont {Cassisi}, \citenamefont {Coppola}, \citenamefont
  {Dall'Ora}, \citenamefont {Gilmozzi}, \citenamefont {Marconi}, \citenamefont
  {Monelli}, \citenamefont {Romaniello}, \citenamefont {Stetson}, \citenamefont
  {Thévenin},\ and\ \citenamefont {Walker}}]{Fabrizio:2016}%
  \BibitemOpen
  \bibfield  {author} {\bibinfo {author} {\bibfnamefont {M.}~\bibnamefont
  {Fabrizio}}, \bibinfo {author} {\bibfnamefont {G.}~\bibnamefont {Bono}},
  \bibinfo {author} {\bibfnamefont {M.}~\bibnamefont {Nonino}}, \bibinfo
  {author} {\bibfnamefont {E.~L.}\ \bibnamefont {Łokas}}, \bibinfo {author}
  {\bibfnamefont {I.}~\bibnamefont {Ferraro}}, \bibinfo {author} {\bibfnamefont
  {G.}~\bibnamefont {Iannicola}}, \bibinfo {author} {\bibfnamefont
  {R.}~\bibnamefont {Buonanno}}, \bibinfo {author} {\bibfnamefont
  {S.}~\bibnamefont {Cassisi}}, \bibinfo {author} {\bibfnamefont
  {G.}~\bibnamefont {Coppola}}, \bibinfo {author} {\bibfnamefont
  {M.}~\bibnamefont {Dall'Ora}}, \bibinfo {author} {\bibfnamefont
  {R.}~\bibnamefont {Gilmozzi}}, \bibinfo {author} {\bibfnamefont
  {M.}~\bibnamefont {Marconi}}, \bibinfo {author} {\bibfnamefont
  {M.}~\bibnamefont {Monelli}}, \bibinfo {author} {\bibfnamefont
  {M.}~\bibnamefont {Romaniello}}, \bibinfo {author} {\bibfnamefont {P.~B.}\
  \bibnamefont {Stetson}}, \bibinfo {author} {\bibfnamefont {F.}~\bibnamefont
  {Thévenin}},\ and\ \bibinfo {author} {\bibfnamefont {A.~R.}\ \bibnamefont
  {Walker}},\ }\bibfield  {title} {\bibinfo {title} {The carina project. x. on
  the kinematics of old and intermediate-age stellar populations1,2.},\ }\href
  {https://doi.org/https://doi.org/10.3847/0004-637X/830/2/126} {\bibfield
  {journal} {\bibinfo  {journal} {Astrophys. J.}\ }\textbf {\bibinfo {volume}
  {830}},\ \bibinfo {pages} {17} (\bibinfo {year} {2016})}\BibitemShut
  {NoStop}%
\bibitem [{\citenamefont {Li}\ \emph {et~al.}(2018)\citenamefont {Li} \emph
  {et~al.}}]{Li:2018}%
  \BibitemOpen
  \bibfield  {author} {\bibinfo {author} {\bibfnamefont {T.~S.}\ \bibnamefont
  {Li}} \emph {et~al.},\ }\bibfield  {title} {\bibinfo {title} {The first
  tidally disrupted ultra-faint dwarf galaxy?: A spectroscopic analysis of the
  tucana iii stream.},\ }\href
  {https://doi.org/https://doi.org/10.3847/1538-4357/aadf91} {\bibfield
  {journal} {\bibinfo  {journal} {Astrophys. J.}\ }\textbf {\bibinfo {volume}
  {866}},\ \bibinfo {pages} {23} (\bibinfo {year} {2018})}\BibitemShut
  {NoStop}%
\bibitem [{\citenamefont {Newby}\ \emph {et~al.}(2013)\citenamefont {Newby},
  \citenamefont {Cole}, \citenamefont {Newberg}, \citenamefont {Desell},
  \citenamefont {Magdon-Ismail}, \citenamefont {Szymanski}, \citenamefont
  {Varela}, \citenamefont {Willett},\ and\ \citenamefont {Yanny}}]{Newby:2013}%
  \BibitemOpen
  \bibfield  {author} {\bibinfo {author} {\bibfnamefont {M.}~\bibnamefont
  {Newby}}, \bibinfo {author} {\bibfnamefont {N.}~\bibnamefont {Cole}},
  \bibinfo {author} {\bibfnamefont {H.~J.}\ \bibnamefont {Newberg}}, \bibinfo
  {author} {\bibfnamefont {T.}~\bibnamefont {Desell}}, \bibinfo {author}
  {\bibfnamefont {M.}~\bibnamefont {Magdon-Ismail}}, \bibinfo {author}
  {\bibfnamefont {B.}~\bibnamefont {Szymanski}}, \bibinfo {author}
  {\bibfnamefont {C.}~\bibnamefont {Varela}}, \bibinfo {author} {\bibfnamefont
  {B.}~\bibnamefont {Willett}},\ and\ \bibinfo {author} {\bibfnamefont
  {B.}~\bibnamefont {Yanny}},\ }\bibfield  {title} {\bibinfo {title} {A spatial
  characterization of the sagittarius dwarf galaxy tidal tails.},\ }\href
  {https://doi.org/https://doi.org/10.1088/0004-6256/145/6/163} {\bibfield
  {journal} {\bibinfo  {journal} {Astronomical. J.}\ }\textbf {\bibinfo
  {volume} {145}},\ \bibinfo {pages} {19} (\bibinfo {year} {2013})}\BibitemShut
  {NoStop}%
\bibitem [{\citenamefont {Widrow}\ and\ \citenamefont
  {Kaiser}(1993)}]{Widrow:1993}%
  \BibitemOpen
  \bibfield  {author} {\bibinfo {author} {\bibfnamefont {L.~M.}\ \bibnamefont
  {Widrow}}\ and\ \bibinfo {author} {\bibfnamefont {N.}~\bibnamefont
  {Kaiser}},\ }\bibfield  {title} {\bibinfo {title} {Using the schroedinger
  equation to simulate collisionless matter.},\ }\href
  {https://doi.org/https://doi.org/10.1086/187073} {\bibfield  {journal}
  {\bibinfo  {journal} {Astrophys. J.}\ }\textbf {\bibinfo {volume} {416}},\
  \bibinfo {pages} {L71} (\bibinfo {year} {1993})}\BibitemShut {NoStop}%
\bibitem [{\citenamefont {Hu}\ \emph {et~al.}(2000)\citenamefont {Hu},
  \citenamefont {Barkana},\ and\ \citenamefont {Gruzinov}}]{Hu:2000}%
  \BibitemOpen
  \bibfield  {author} {\bibinfo {author} {\bibfnamefont {W.}~\bibnamefont
  {Hu}}, \bibinfo {author} {\bibfnamefont {R.}~\bibnamefont {Barkana}},\ and\
  \bibinfo {author} {\bibfnamefont {A.}~\bibnamefont {Gruzinov}},\ }\bibfield
  {title} {\bibinfo {title} {Fuzzy cold dark matter: The wave properties of
  ultralight particles.},\ }\href
  {https://doi.org/https://doi.org/10.1103/PhysRevLett.85.1158} {\bibfield
  {journal} {\bibinfo  {journal} {Phys. Rev. Lett.}\ }\textbf {\bibinfo
  {volume} {85}},\ \bibinfo {pages} {1158} (\bibinfo {year}
  {2000})}\BibitemShut {NoStop}%
\bibitem [{\citenamefont {Schwabe}\ \emph {et~al.}(2016)\citenamefont
  {Schwabe}, \citenamefont {Niemeyer},\ and\ \citenamefont
  {Engels}}]{Schwabe:2016}%
  \BibitemOpen
  \bibfield  {author} {\bibinfo {author} {\bibfnamefont {B.}~\bibnamefont
  {Schwabe}}, \bibinfo {author} {\bibfnamefont {J.~C.}\ \bibnamefont
  {Niemeyer}},\ and\ \bibinfo {author} {\bibfnamefont {J.~F.}\ \bibnamefont
  {Engels}},\ }\bibfield  {title} {\bibinfo {title} {Simulations of solitonic
  core mergers in ultralight axion dark matter cosmologies.},\ }\bibfield
  {journal} {\bibinfo  {journal} {Phys. Rev. D.}\ }\textbf {\bibinfo {volume}
  {94}},\ \href {https://doi.org/https://doi.org/ 10.1103/PhysRevD.94.043513}
  {https://doi.org/ 10.1103/PhysRevD.94.043513} (\bibinfo {year}
  {2016})\BibitemShut {NoStop}%
\bibitem [{\citenamefont {Mocz}\ \emph {et~al.}(2017)\citenamefont {Mocz},
  \citenamefont {Vogelsberger}, \citenamefont {Robles}, \citenamefont {Zavala},
  \citenamefont {Boylan-Kolchin}, \citenamefont {Fialkov},\ and\ \citenamefont
  {Hernquist}}]{Mocz:2017}%
  \BibitemOpen
  \bibfield  {author} {\bibinfo {author} {\bibfnamefont {P.}~\bibnamefont
  {Mocz}}, \bibinfo {author} {\bibfnamefont {M.}~\bibnamefont {Vogelsberger}},
  \bibinfo {author} {\bibfnamefont {V.~H.}\ \bibnamefont {Robles}}, \bibinfo
  {author} {\bibfnamefont {J.}~\bibnamefont {Zavala}}, \bibinfo {author}
  {\bibfnamefont {M.}~\bibnamefont {Boylan-Kolchin}}, \bibinfo {author}
  {\bibfnamefont {A.}~\bibnamefont {Fialkov}},\ and\ \bibinfo {author}
  {\bibfnamefont {L.}~\bibnamefont {Hernquist}},\ }\bibfield  {title} {\bibinfo
  {title} {Galaxy formation with becdm - i. turbulence and relaxation of
  idealized haloes.},\ }\href
  {https://doi.org/https://doi.org/10.1093/mnras/stx1887} {\bibfield  {journal}
  {\bibinfo  {journal} {Mon. Not. R. Astron. Soc.}\ }\textbf {\bibinfo {volume}
  {471}},\ \bibinfo {pages} {4559} (\bibinfo {year} {2017})}\BibitemShut
  {NoStop}%
\bibitem [{\citenamefont {May}\ and\ \citenamefont
  {Springel}(2021)}]{May:2021}%
  \BibitemOpen
  \bibfield  {author} {\bibinfo {author} {\bibfnamefont {S.}~\bibnamefont
  {May}}\ and\ \bibinfo {author} {\bibfnamefont {V.}~\bibnamefont {Springel}},\
  }\bibfield  {title} {\bibinfo {title} {Structure formation in large-volume
  cosmological simulations of fuzzy dark matter: Impact of the non-linear
  dynamics.},\ }\bibfield  {journal} {\bibinfo  {journal} {Mon. Not. R. Astron.
  Soc.}\ }\textbf {\bibinfo {volume} {Advanced access}},\ \href
  {https://doi.org/https://doi.org/10.1093/mnras/stab1764}
  {https://doi.org/10.1093/mnras/stab1764} (\bibinfo {year} {2021})\BibitemShut
  {NoStop}%
\bibitem [{\citenamefont {Navarro}\ and\ \citenamefont
  {Frenk}(1996)}]{Navarro:1996}%
  \BibitemOpen
  \bibfield  {author} {\bibinfo {author} {\bibfnamefont {J.}~\bibnamefont
  {Navarro}}\ and\ \bibinfo {author} {\bibfnamefont {C.~S.}\ \bibnamefont
  {Frenk}},\ }\bibfield  {title} {\bibinfo {title} {The structure of cold dark
  matter halos.},\ }\href {https://doi.org/https://doi.org/10.1086/177173}
  {\bibfield  {journal} {\bibinfo  {journal} {Astrophys. J.}\ }\textbf
  {\bibinfo {volume} {462}},\ \bibinfo {pages} {563} (\bibinfo {year}
  {1996})}\BibitemShut {NoStop}%
\bibitem [{\citenamefont {Woo}\ and\ \citenamefont {Chiueh}(2009)}]{Woo:2009}%
  \BibitemOpen
  \bibfield  {author} {\bibinfo {author} {\bibfnamefont {T.~P.}\ \bibnamefont
  {Woo}}\ and\ \bibinfo {author} {\bibfnamefont {T.}~\bibnamefont {Chiueh}},\
  }\bibfield  {title} {\bibinfo {title} {High-resolution simulation on
  structure formation with extremely light bosonic dark matter.},\ }\href
  {https://doi.org/https://doi.org/ 10.1088/0004-637X/697/1/850} {\bibfield
  {journal} {\bibinfo  {journal} {Astrophys. J.}\ }\textbf {\bibinfo {volume}
  {697}},\ \bibinfo {pages} {850} (\bibinfo {year} {2009})}\BibitemShut
  {NoStop}%
\bibitem [{\citenamefont {Bryan}\ and\ \citenamefont
  {Norman}(1998)}]{Bryan:1998}%
  \BibitemOpen
  \bibfield  {author} {\bibinfo {author} {\bibfnamefont {G.~L.}\ \bibnamefont
  {Bryan}}\ and\ \bibinfo {author} {\bibfnamefont {M.~L.}\ \bibnamefont
  {Norman}},\ }\bibfield  {title} {\bibinfo {title} {Statistical properties of
  x-ray clusters: Analytic and numerical comparisons.},\ }\href
  {https://doi.org/https://doi.org/10.48550/arXiv.astro-ph/9710107} {\bibfield
  {journal} {\bibinfo  {journal} {Astrophys. J.}\ }\textbf {\bibinfo {volume}
  {495}},\ \bibinfo {pages} {80} (\bibinfo {year} {1998})}\BibitemShut
  {NoStop}%
\bibitem [{\citenamefont {Binney}\ and\ \citenamefont
  {Tremaine}(2008)}]{Binney:2008}%
  \BibitemOpen
  \bibfield  {author} {\bibinfo {author} {\bibfnamefont {J.}~\bibnamefont
  {Binney}}\ and\ \bibinfo {author} {\bibfnamefont {S.}~\bibnamefont
  {Tremaine}},\ }\bibfield  {title} {\bibinfo {title} {Galactic dynamics:
  Second edition},\ }\href@noop {} {\bibfield  {journal} {\bibinfo  {journal}
  {Princeton Univ.}\ } (\bibinfo {year} {2008})}\BibitemShut {NoStop}%
\bibitem [{\citenamefont {Fraternali}\ \emph {et~al.}(2009)\citenamefont
  {Fraternali}, \citenamefont {Tolstoy}, \citenamefont {Irwin},\ and\
  \citenamefont {Cole}}]{Fraternali:2009}%
  \BibitemOpen
  \bibfield  {author} {\bibinfo {author} {\bibfnamefont {F.}~\bibnamefont
  {Fraternali}}, \bibinfo {author} {\bibfnamefont {E.}~\bibnamefont {Tolstoy}},
  \bibinfo {author} {\bibfnamefont {M.~J.}\ \bibnamefont {Irwin}},\ and\
  \bibinfo {author} {\bibfnamefont {A.~A.}\ \bibnamefont {Cole}},\ }\bibfield
  {title} {\bibinfo {title} {Life at the periphery of the local group: the
  kinematics of the tucana dwarf galaxy.},\ }\href
  {https://doi.org/https://doi.org/10.1051/0004-6361/200810830} {\bibfield
  {journal} {\bibinfo  {journal} {Astron. Astrophys.}\ }\textbf {\bibinfo
  {volume} {499}},\ \bibinfo {pages} {121} (\bibinfo {year}
  {2009})}\BibitemShut {NoStop}%
\bibitem [{\citenamefont {Taibi}\ \emph {et~al.}(2018)\citenamefont {Taibi},
  \citenamefont {Battaglia}, \citenamefont {Kacharov}, \citenamefont {Rejkuba},
  \citenamefont {Irwin}, \citenamefont {Leaman}, \citenamefont {Zoccali},
  \citenamefont {Tolstoy},\ and\ \citenamefont {Jablonka}}]{Taibi:2018}%
  \BibitemOpen
  \bibfield  {author} {\bibinfo {author} {\bibfnamefont {S.}~\bibnamefont
  {Taibi}}, \bibinfo {author} {\bibfnamefont {G.}~\bibnamefont {Battaglia}},
  \bibinfo {author} {\bibfnamefont {N.}~\bibnamefont {Kacharov}}, \bibinfo
  {author} {\bibfnamefont {M.}~\bibnamefont {Rejkuba}}, \bibinfo {author}
  {\bibfnamefont {M.}~\bibnamefont {Irwin}}, \bibinfo {author} {\bibfnamefont
  {R.}~\bibnamefont {Leaman}}, \bibinfo {author} {\bibfnamefont
  {M.}~\bibnamefont {Zoccali}}, \bibinfo {author} {\bibfnamefont
  {E.}~\bibnamefont {Tolstoy}},\ and\ \bibinfo {author} {\bibfnamefont
  {P.}~\bibnamefont {Jablonka}},\ }\bibfield  {title} {\bibinfo {title}
  {Stellar chemo-kinematics of the cetus dwarf spheroidal galaxy.},\ }\href
  {https://doi.org/https://doi.org/10.1051/0004-6361/201833414} {\bibfield
  {journal} {\bibinfo  {journal} {Astron. Astrophys.}\ }\textbf {\bibinfo
  {volume} {618}},\ \bibinfo {pages} {22} (\bibinfo {year} {2018})}\BibitemShut
  {NoStop}%
\bibitem [{\citenamefont {Kirby}\ \emph {et~al.}(2017)\citenamefont {Kirby},
  \citenamefont {Rizzi}, \citenamefont {Held}, \citenamefont {Cohen},
  \citenamefont {Cole}, \citenamefont {Manning}, \citenamefont {Skillman},\
  and\ \citenamefont {Weisz}}]{Kirby:2017}%
  \BibitemOpen
  \bibfield  {author} {\bibinfo {author} {\bibfnamefont {E.~N.}\ \bibnamefont
  {Kirby}}, \bibinfo {author} {\bibfnamefont {L.}~\bibnamefont {Rizzi}},
  \bibinfo {author} {\bibfnamefont {E.~V.}\ \bibnamefont {Held}}, \bibinfo
  {author} {\bibfnamefont {J.~G.}\ \bibnamefont {Cohen}}, \bibinfo {author}
  {\bibfnamefont {A.~A.}\ \bibnamefont {Cole}}, \bibinfo {author}
  {\bibfnamefont {E.~M.}\ \bibnamefont {Manning}}, \bibinfo {author}
  {\bibfnamefont {E.~D.}\ \bibnamefont {Skillman}},\ and\ \bibinfo {author}
  {\bibfnamefont {D.~R.}\ \bibnamefont {Weisz}},\ }\bibfield  {title} {\bibinfo
  {title} {A chemistry and kinematics of the late-forming dwarf irregular
  galaxies leo a, aquarius, and sagittarius dig},\ }\href
  {https://doi.org/https://doi.org/10.3847/1538-4357/834/1/9} {\bibfield
  {journal} {\bibinfo  {journal} {Astrophys. J.}\ }\textbf {\bibinfo {volume}
  {834}},\ \bibinfo {pages} {19} (\bibinfo {year} {2017})}\BibitemShut
  {NoStop}%
\bibitem [{\citenamefont {Schive}\ \emph {et~al.}(2020)\citenamefont {Schive},
  \citenamefont {Chiueh},\ and\ \citenamefont {Broadhurst}}]{Schive:2020}%
  \BibitemOpen
  \bibfield  {author} {\bibinfo {author} {\bibfnamefont {H.-Y.}\ \bibnamefont
  {Schive}}, \bibinfo {author} {\bibfnamefont {T.}~\bibnamefont {Chiueh}},\
  and\ \bibinfo {author} {\bibfnamefont {T.}~\bibnamefont {Broadhurst}},\
  }\bibfield  {title} {\bibinfo {title} {Soliton random walk and the
  cluster-stripping problem in ultralight dark matter.},\ }\bibfield  {journal}
  {\bibinfo  {journal} {Phys. Rev. Lett.}\ }\textbf {\bibinfo {volume} {124}},\
  \href {https://doi.org/https://doi.org/10.1103/PhysRevLett.124.201301}
  {https://doi.org/10.1103/PhysRevLett.124.201301} (\bibinfo {year}
  {2020})\BibitemShut {NoStop}%
\bibitem [{\citenamefont {Irwin}\ and\ \citenamefont
  {Hatzidimitriou}(1995)}]{Irwin:1995}%
  \BibitemOpen
  \bibfield  {author} {\bibinfo {author} {\bibfnamefont {M.}~\bibnamefont
  {Irwin}}\ and\ \bibinfo {author} {\bibfnamefont {D.}~\bibnamefont
  {Hatzidimitriou}},\ }\bibfield  {title} {\bibinfo {title} {Structural
  parameters for the galactic dwarf spheroidals.},\ }\href
  {https://doi.org/https://doi.org/10.1093/mnras/277.4.1354} {\bibfield
  {journal} {\bibinfo  {journal} {Mon. Not. R. Astron. Soc.}\ }\textbf
  {\bibinfo {volume} {277}},\ \bibinfo {pages} {1354} (\bibinfo {year}
  {1995})}\BibitemShut {NoStop}%
\bibitem [{\citenamefont {Sohn}\ \emph {et~al.}(2007)\citenamefont {Sohn},
  \citenamefont {Majewski}, \citenamefont {Muñoz}, \citenamefont {Kunkel},
  \citenamefont {Johnston}, \citenamefont {Ostheimer}, \citenamefont
  {Guhathakurta}, \citenamefont {Patterson}, \citenamefont {Siegel},\ and\
  \citenamefont {Cooper}}]{Sohn:2007}%
  \BibitemOpen
  \bibfield  {author} {\bibinfo {author} {\bibfnamefont {S.~T.}\ \bibnamefont
  {Sohn}}, \bibinfo {author} {\bibfnamefont {S.~R.}\ \bibnamefont {Majewski}},
  \bibinfo {author} {\bibfnamefont {R.~R.}\ \bibnamefont {Muñoz}}, \bibinfo
  {author} {\bibfnamefont {W.~E.}\ \bibnamefont {Kunkel}}, \bibinfo {author}
  {\bibfnamefont {K.~V.}\ \bibnamefont {Johnston}}, \bibinfo {author}
  {\bibfnamefont {J.~C.}\ \bibnamefont {Ostheimer}}, \bibinfo {author}
  {\bibfnamefont {P.}~\bibnamefont {Guhathakurta}}, \bibinfo {author}
  {\bibfnamefont {R.~J.}\ \bibnamefont {Patterson}}, \bibinfo {author}
  {\bibfnamefont {M.~H.}\ \bibnamefont {Siegel}},\ and\ \bibinfo {author}
  {\bibfnamefont {M.~C.}\ \bibnamefont {Cooper}},\ }\bibfield  {title}
  {\bibinfo {title} {Exploring halo substructure with giant stars. x. extended
  dark matter or tidal disruption?: The case for the leo i dwarf spheroidal
  galaxy.},\ }\href {https://doi.org/https://doi.org/10.1086/518302} {\bibfield
   {journal} {\bibinfo  {journal} {Astrophys. J.}\ }\textbf {\bibinfo {volume}
  {663}},\ \bibinfo {pages} {960} (\bibinfo {year} {2007})}\BibitemShut
  {NoStop}%
\bibitem [{\citenamefont {Martínez-Delgado}\ \emph {et~al.}(2001)\citenamefont
  {Martínez-Delgado}, \citenamefont {Alonso-García}, \citenamefont
  {Aparicio},\ and\ \citenamefont {Gómez-Flechoso}}]{Martinez-Delgado:2001}%
  \BibitemOpen
  \bibfield  {author} {\bibinfo {author} {\bibfnamefont {D.}~\bibnamefont
  {Martínez-Delgado}}, \bibinfo {author} {\bibfnamefont {J.}~\bibnamefont
  {Alonso-García}}, \bibinfo {author} {\bibfnamefont {A.}~\bibnamefont
  {Aparicio}},\ and\ \bibinfo {author} {\bibfnamefont {M.~A.}\ \bibnamefont
  {Gómez-Flechoso}},\ }\bibfield  {title} {\bibinfo {title} {A tidal extension
  in the ursa minor dwarf spheroidal galaxy.},\ }\href
  {https://doi.org/https://doi.org/10.1086/319150} {\bibfield  {journal}
  {\bibinfo  {journal} {Astrophys. J.}\ }\textbf {\bibinfo {volume} {549}},\
  \bibinfo {pages} {L63} (\bibinfo {year} {2001})}\BibitemShut {NoStop}%
\bibitem [{\citenamefont {Walker}\ and\ \citenamefont
  {Peñarrubia}(2011)}]{Walker:2011}%
  \BibitemOpen
  \bibfield  {author} {\bibinfo {author} {\bibfnamefont {M.~G.}\ \bibnamefont
  {Walker}}\ and\ \bibinfo {author} {\bibfnamefont {J.}~\bibnamefont
  {Peñarrubia}},\ }\bibfield  {title} {\bibinfo {title} {A method for
  measuring (slopes of) the mass profiles of dwarf spheroidal galaxies.},\
  }\href {https://doi.org/https://doi.org/10.1088/0004-637X/742/1/20}
  {\bibfield  {journal} {\bibinfo  {journal} {Astrophys. J.}\ }\textbf
  {\bibinfo {volume} {742}},\ \bibinfo {pages} {19} (\bibinfo {year}
  {2011})}\BibitemShut {NoStop}%
\bibitem [{\citenamefont {Amorisco}\ and\ \citenamefont
  {Evans}(2012)}]{Amorisco:2012}%
  \BibitemOpen
  \bibfield  {author} {\bibinfo {author} {\bibfnamefont {N.~C.}\ \bibnamefont
  {Amorisco}}\ and\ \bibinfo {author} {\bibfnamefont {N.~W.}\ \bibnamefont
  {Evans}},\ }\bibfield  {title} {\bibinfo {title} {Dark matter cores and
  cusps: the case of multiple stellar populations in dwarf spheroidals.},\
  }\href {https://doi.org/https://doi.org/10.1111/j.1365-2966.2011.19684.x}
  {\bibfield  {journal} {\bibinfo  {journal} {Mon. Not. R. Astron. Soc.}\
  }\textbf {\bibinfo {volume} {419}},\ \bibinfo {pages} {184} (\bibinfo {year}
  {2012})}\BibitemShut {NoStop}%
\bibitem [{\citenamefont {Battaglia}\ \emph
  {et~al.}(2012{\natexlab{a}})\citenamefont {Battaglia}, \citenamefont
  {Rejkuba}, \citenamefont {Tolstoy}, \citenamefont {Irwin},\ and\
  \citenamefont {Beccari}}]{Battaglia:20122}%
  \BibitemOpen
  \bibfield  {author} {\bibinfo {author} {\bibfnamefont {G.}~\bibnamefont
  {Battaglia}}, \bibinfo {author} {\bibfnamefont {M.}~\bibnamefont {Rejkuba}},
  \bibinfo {author} {\bibfnamefont {E.}~\bibnamefont {Tolstoy}}, \bibinfo
  {author} {\bibfnamefont {M.~J.}\ \bibnamefont {Irwin}},\ and\ \bibinfo
  {author} {\bibfnamefont {G.~A.}\ \bibnamefont {Beccari}},\ }\bibfield
  {title} {\bibinfo {title} {The extensive age gradient of the carina dwarf
  galaxy.},\ }\href
  {https://doi.org/https://doi.org/10.1088/2041-8205/761/2/L31} {\bibfield
  {journal} {\bibinfo  {journal} {Astrophys. J.}\ }\textbf {\bibinfo {volume}
  {761}},\ \bibinfo {pages} {6} (\bibinfo {year}
  {2012}{\natexlab{a}})}\BibitemShut {NoStop}%
\bibitem [{\citenamefont {Niemeyer}(2020)}]{Niemeyer:2020}%
  \BibitemOpen
  \bibfield  {author} {\bibinfo {author} {\bibfnamefont {J.~C.}\ \bibnamefont
  {Niemeyer}},\ }\bibfield  {title} {\bibinfo {title} {Small-scale structure of
  fuzzy and axion-like dark matter.},\ }\bibfield  {journal} {\bibinfo
  {journal} {P. in Particle $\&$ Nuclear Phys.}\ }\textbf {\bibinfo {volume}
  {113}},\ \href {https://doi.org/https://doi.org/10.1016/j.ppnp.2020.103787}
  {https://doi.org/10.1016/j.ppnp.2020.103787} (\bibinfo {year}
  {2020})\BibitemShut {NoStop}%
\bibitem [{\citenamefont {Veltmaat}\ \emph {et~al.}(2020)\citenamefont
  {Veltmaat}, \citenamefont {Schwabe},\ and\ \citenamefont
  {Niemeyer}}]{Veltmaat:2019}%
  \BibitemOpen
  \bibfield  {author} {\bibinfo {author} {\bibfnamefont {J.}~\bibnamefont
  {Veltmaat}}, \bibinfo {author} {\bibfnamefont {B.}~\bibnamefont {Schwabe}},\
  and\ \bibinfo {author} {\bibfnamefont {J.~C.}\ \bibnamefont {Niemeyer}},\
  }\bibfield  {title} {\bibinfo {title} {Baryon-driven growth of solitonic
  cores in fuzzy dark matter halos.},\ }\bibfield  {journal} {\bibinfo
  {journal} {Phys. Rev. Lett.}\ }\textbf {\bibinfo {volume} {101}},\ \href
  {https://doi.org/https://doi.org/ 10.1103/PhysRevD.101.083518}
  {https://doi.org/ 10.1103/PhysRevD.101.083518} (\bibinfo {year}
  {2020})\BibitemShut {NoStop}%
\bibitem [{\citenamefont {Hui}\ \emph {et~al.}(2021)\citenamefont {Hui},
  \citenamefont {Joyce},\ and\ \citenamefont {Landry}}]{Hui:2020}%
  \BibitemOpen
  \bibfield  {author} {\bibinfo {author} {\bibfnamefont {L.}~\bibnamefont
  {Hui}}, \bibinfo {author} {\bibfnamefont {A.}~\bibnamefont {Joyce}},\ and\
  \bibinfo {author} {\bibfnamefont {X.}~\bibnamefont {Landry}, \bibfnamefont
  {M.~J.and~Li}},\ }\bibfield  {title} {\bibinfo {title} {Vortices and waves in
  light dark matter},\ }\href
  {https://doi.org/https://doi.org/10.1088/1475-7516/2021/01/011} {\bibfield
  {journal} {\bibinfo  {journal} {Journal of Cosmology and Astroparticle
  Physics}\ }\textbf {\bibinfo {volume} {011}}}\BibitemShut {NoStop}%
\bibitem [{\citenamefont {Chen}\ \emph {et~al.}(2017)\citenamefont {Chen},
  \citenamefont {Schive},\ and\ \citenamefont {Chiueh}}]{Chen:2017}%
  \BibitemOpen
  \bibfield  {author} {\bibinfo {author} {\bibfnamefont {S.-R.}\ \bibnamefont
  {Chen}}, \bibinfo {author} {\bibfnamefont {H.-Y.}\ \bibnamefont {Schive}},\
  and\ \bibinfo {author} {\bibfnamefont {T.}~\bibnamefont {Chiueh}},\
  }\bibfield  {title} {\bibinfo {title} {Jeans analysis for dwarf spheroidal
  galaxies in wave dark matter.},\ }\href
  {https://doi.org/https://doi.org/10.1093/mnras/stx449} {\bibfield  {journal}
  {\bibinfo  {journal} {Mon. Not. R. Astron. Soc.}\ }\textbf {\bibinfo {volume}
  {468}},\ \bibinfo {pages} {1338} (\bibinfo {year} {2017})}\BibitemShut
  {NoStop}%
\bibitem [{\citenamefont {González-Morales}\ \emph {et~al.}(2017)\citenamefont
  {González-Morales} \emph {et~al.}}]{GonzalesMorales:2017}%
  \BibitemOpen
  \bibfield  {author} {\bibinfo {author} {\bibfnamefont {A.~X.}\ \bibnamefont
  {González-Morales}} \emph {et~al.},\ }\bibfield  {title} {\bibinfo {title}
  {Population gradient in the sextans dsph: comprehensive mapping of a dwarf
  galaxy by suprime-cam.},\ }\href
  {https://doi.org/https://doi.org/10.1093/mnras/stx1941} {\bibfield  {journal}
  {\bibinfo  {journal} {Mon. Not. R. Astron. Soc.}\ }\textbf {\bibinfo {volume}
  {472}},\ \bibinfo {pages} {1346} (\bibinfo {year} {2017})}\BibitemShut
  {NoStop}%
\bibitem [{\citenamefont {Nori}\ \emph {et~al.}(2023)\citenamefont {Nori},
  \citenamefont {Macciò},\ and\ \citenamefont {Baldi}}]{Nori:2023}%
  \BibitemOpen
  \bibfield  {author} {\bibinfo {author} {\bibfnamefont {M.}~\bibnamefont
  {Nori}}, \bibinfo {author} {\bibfnamefont {A.~V.}\ \bibnamefont {Macciò}},\
  and\ \bibinfo {author} {\bibfnamefont {M.}~\bibnamefont {Baldi}},\ }\bibfield
   {title} {\bibinfo {title} {Fuzzy aquarius: evolution of a milky-way like
  system in the fuzzy dark matter scenario.},\ }\href
  {https://doi.org/https://doi.org/10.1093/mnras/stad1081} {\bibfield
  {journal} {\bibinfo  {journal} {Mon. Not. R. Astron. Soc.}\ }\textbf
  {\bibinfo {volume} {522}},\ \bibinfo {pages} {1451} (\bibinfo {year}
  {2023})}\BibitemShut {NoStop}%
\bibitem [{\citenamefont {Roderick}\ \emph {et~al.}(2016)\citenamefont
  {Roderick}, \citenamefont {Jerjen}, \citenamefont {Da~Costa},\ and\
  \citenamefont {Mackey}}]{Roderick:2016}%
  \BibitemOpen
  \bibfield  {author} {\bibinfo {author} {\bibfnamefont {T.~A.}\ \bibnamefont
  {Roderick}}, \bibinfo {author} {\bibfnamefont {H.}~\bibnamefont {Jerjen}},
  \bibinfo {author} {\bibfnamefont {G.~S.}\ \bibnamefont {Da~Costa}},\ and\
  \bibinfo {author} {\bibfnamefont {A.~D.}\ \bibnamefont {Mackey}},\ }\bibfield
   {title} {\bibinfo {title} {Structural analysis of the sextans dwarf
  spheroidal galaxy.},\ }\href
  {https://doi.org/https://doi.org/10.1093/mnras/stw949} {\bibfield  {journal}
  {\bibinfo  {journal} {Mon. Not. R. Astron. Soc.}\ }\textbf {\bibinfo {volume}
  {460}},\ \bibinfo {pages} {30} (\bibinfo {year} {2016})}\BibitemShut
  {NoStop}%
\bibitem [{\citenamefont {Okamoto}\ \emph {et~al.}(2017)\citenamefont
  {Okamoto}, \citenamefont {Arimoto}, \citenamefont {Tolstoy}, \citenamefont
  {Jablonka}, \citenamefont {Irwin}, \citenamefont {Komiyama}, \citenamefont
  {Yamada},\ and\ \citenamefont {Onodera}}]{Okamoto:2017}%
  \BibitemOpen
  \bibfield  {author} {\bibinfo {author} {\bibfnamefont {S.}~\bibnamefont
  {Okamoto}}, \bibinfo {author} {\bibfnamefont {N.}~\bibnamefont {Arimoto}},
  \bibinfo {author} {\bibfnamefont {E.}~\bibnamefont {Tolstoy}}, \bibinfo
  {author} {\bibfnamefont {P.}~\bibnamefont {Jablonka}}, \bibinfo {author}
  {\bibfnamefont {M.~J.}\ \bibnamefont {Irwin}}, \bibinfo {author}
  {\bibfnamefont {Y.}~\bibnamefont {Komiyama}}, \bibinfo {author}
  {\bibfnamefont {Y.}~\bibnamefont {Yamada}},\ and\ \bibinfo {author}
  {\bibfnamefont {M.}~\bibnamefont {Onodera}},\ }\bibfield  {title} {\bibinfo
  {title} {Population gradient in the sextans dsph: comprehensive mapping of a
  dwarf galaxy by suprime-cam.},\ }\href
  {https://doi.org/https://doi.org/10.1093/mnras/stx086} {\bibfield  {journal}
  {\bibinfo  {journal} {Mon. Not. R. Astron. Soc.}\ }\textbf {\bibinfo {volume}
  {467}},\ \bibinfo {pages} {208} (\bibinfo {year} {2017})}\BibitemShut
  {NoStop}%
\bibitem [{\citenamefont {Chan}\ \emph {et~al.}(2020)\citenamefont {Chan},
  \citenamefont {Schive}, \citenamefont {Wong}, \citenamefont {Chiueh},\ and\
  \citenamefont {Broadhurst}}]{Chan:2020}%
  \BibitemOpen
  \bibfield  {author} {\bibinfo {author} {\bibfnamefont {J.~H.~H.}\
  \bibnamefont {Chan}}, \bibinfo {author} {\bibfnamefont {H.-Y.}\ \bibnamefont
  {Schive}}, \bibinfo {author} {\bibfnamefont {S.-K.}\ \bibnamefont {Wong}},
  \bibinfo {author} {\bibfnamefont {T.}~\bibnamefont {Chiueh}},\ and\ \bibinfo
  {author} {\bibfnamefont {T.}~\bibnamefont {Broadhurst}},\ }\bibfield  {title}
  {\bibinfo {title} {Multiple images and flux ratio anomaly of fuzzy
  gravitational lenses.},\ }\bibfield  {journal} {\bibinfo  {journal} {ArXiv}\
  }\href {https://doi.org/https://doi.org/10.1103/PhysRevLett.125.111102}
  {https://doi.org/10.1103/PhysRevLett.125.111102} (\bibinfo {year}
  {2020})\BibitemShut {NoStop}%
\bibitem [{\citenamefont {Broadhurst}\ \emph {et~al.}(2020)\citenamefont
  {Broadhurst}, \citenamefont {de~Martino}, \citenamefont {Luu}, \citenamefont
  {Smoot},\ and\ \citenamefont {Tye}}]{Broadhurst:2020}%
  \BibitemOpen
  \bibfield  {author} {\bibinfo {author} {\bibfnamefont {T.}~\bibnamefont
  {Broadhurst}}, \bibinfo {author} {\bibfnamefont {I.}~\bibnamefont
  {de~Martino}}, \bibinfo {author} {\bibfnamefont {H.~N.}\ \bibnamefont {Luu}},
  \bibinfo {author} {\bibfnamefont {G.~F.}\ \bibnamefont {Smoot}},\ and\
  \bibinfo {author} {\bibfnamefont {S.~H.~H.}\ \bibnamefont {Tye}},\ }\bibfield
   {title} {\bibinfo {title} {Ghostly galaxies as solitons of bose-einstein
  dark matter.},\ }\bibfield  {journal} {\bibinfo  {journal} {Phys. Rev. D.}\
  }\textbf {\bibinfo {volume} {101}},\ \href
  {https://doi.org/https://doi.org/10.1103/PhysRevD.101.083012}
  {https://doi.org/10.1103/PhysRevD.101.083012} (\bibinfo {year}
  {2020})\BibitemShut {NoStop}%
\bibitem [{\citenamefont {Pozo}\ \emph {et~al.}(2020)\citenamefont {Pozo},
  \citenamefont {Broadhurst}, \citenamefont {de~Martino}, \citenamefont {Luu},
  \citenamefont {Smoot}, \citenamefont {Lim},\ and\ \citenamefont
  {Neyrinck}}]{Pozo:2020}%
  \BibitemOpen
  \bibfield  {author} {\bibinfo {author} {\bibfnamefont {A.}~\bibnamefont
  {Pozo}}, \bibinfo {author} {\bibfnamefont {T.}~\bibnamefont {Broadhurst}},
  \bibinfo {author} {\bibfnamefont {I.}~\bibnamefont {de~Martino}}, \bibinfo
  {author} {\bibfnamefont {H.~N.}\ \bibnamefont {Luu}}, \bibinfo {author}
  {\bibfnamefont {G.~F.}\ \bibnamefont {Smoot}}, \bibinfo {author}
  {\bibfnamefont {J.}~\bibnamefont {Lim}},\ and\ \bibinfo {author}
  {\bibfnamefont {M.}~\bibnamefont {Neyrinck}},\ }\bibfield  {title} {\bibinfo
  {title} {Wave dark matter and ultra diffuse galaxies.},\ }\href
  {https://doi.org/https://doi.org/10.1093/mnras/stab855} {\bibfield  {journal}
  {\bibinfo  {journal} {Mon. Not. R. Astron. Soc.}\ }\textbf {\bibinfo {volume}
  {504}},\ \bibinfo {pages} {2868} (\bibinfo {year} {2020})}\BibitemShut
  {NoStop}%
\bibitem [{\citenamefont {Bar}\ \emph {et~al.}(2019)\citenamefont {Bar},
  \citenamefont {Blum}, \citenamefont {Eby},\ and\ \citenamefont
  {Sato}}]{Bar:2019}%
  \BibitemOpen
  \bibfield  {author} {\bibinfo {author} {\bibfnamefont {N.}~\bibnamefont
  {Bar}}, \bibinfo {author} {\bibfnamefont {K.}~\bibnamefont {Blum}}, \bibinfo
  {author} {\bibfnamefont {J.}~\bibnamefont {Eby}},\ and\ \bibinfo {author}
  {\bibfnamefont {R.}~\bibnamefont {Sato}},\ }\bibfield  {title} {\bibinfo
  {title} {Ultralight dark matter in disk galaxies},\ }\bibfield  {journal}
  {\bibinfo  {journal} {Phys. Rev. D.}\ }\textbf {\bibinfo {volume} {99}},\
  \href {https://doi.org/https://doi.org/10.1103/PhysRevD.99.103020}
  {https://doi.org/10.1103/PhysRevD.99.103020} (\bibinfo {year}
  {2019})\BibitemShut {NoStop}%
\bibitem [{\citenamefont {Oman}\ \emph {et~al.}(2019)\citenamefont {Oman},
  \citenamefont {Marasco}, \citenamefont {Navarro}, \citenamefont {Frenk},
  \citenamefont {Schaye},\ and\ \citenamefont {Benítez-Llambay}}]{Oman:2019}%
  \BibitemOpen
  \bibfield  {author} {\bibinfo {author} {\bibfnamefont {K.~A.}\ \bibnamefont
  {Oman}}, \bibinfo {author} {\bibfnamefont {A.}~\bibnamefont {Marasco}},
  \bibinfo {author} {\bibfnamefont {J.~F.}\ \bibnamefont {Navarro}}, \bibinfo
  {author} {\bibfnamefont {C.~S.}\ \bibnamefont {Frenk}}, \bibinfo {author}
  {\bibfnamefont {J.}~\bibnamefont {Schaye}},\ and\ \bibinfo {author}
  {\bibfnamefont {A.}~\bibnamefont {Benítez-Llambay}},\ }\bibfield  {title}
  {\bibinfo {title} {Non-circular motions and the diversity of dwarf galaxy
  rotation curves.},\ }\href
  {https://doi.org/https://doi.org/10.1093/mnras/sty2687} {\bibfield  {journal}
  {\bibinfo  {journal} {Mon. Not. R. Astron. Soc.}\ }\textbf {\bibinfo {volume}
  {482}},\ \bibinfo {pages} {821} (\bibinfo {year} {2019})}\BibitemShut
  {NoStop}%
\bibitem [{\citenamefont {Deng}(2018)}]{Deng:2018}%
  \BibitemOpen
  \bibfield  {author} {\bibinfo {author} {\bibfnamefont {H.~o.}\ \bibnamefont
  {Deng}},\ }\bibfield  {title} {\bibinfo {title} {Can light dark matter solve
  the core-cusp problem?},\ }\bibfield  {journal} {\bibinfo  {journal} {Phys.
  Rev. D.}\ }\textbf {\bibinfo {volume} {98}},\ \href
  {https://doi.org/https://doi.org/10.1103/PhysRevD.98.023513}
  {https://doi.org/10.1103/PhysRevD.98.023513} (\bibinfo {year}
  {2018})\BibitemShut {NoStop}%
\bibitem [{\citenamefont {Dutta~Chowdhury}\ \emph {et~al.}(2021)\citenamefont
  {Dutta~Chowdhury}, \citenamefont {van~den Bosch}, \citenamefont {Robles},
  \citenamefont {van Dokkum}, \citenamefont {Chiueh},\ and\ \citenamefont
  {Broadhurst}}]{Dutta:2021}%
  \BibitemOpen
  \bibfield  {author} {\bibinfo {author} {\bibfnamefont {D.}~\bibnamefont
  {Dutta~Chowdhury}}, \bibinfo {author} {\bibfnamefont {F.~C.}\ \bibnamefont
  {van~den Bosch}}, \bibinfo {author} {\bibfnamefont {V.~H.}\ \bibnamefont
  {Robles}}, \bibinfo {author} {\bibfnamefont {H.-Y.}\ \bibnamefont {van
  Dokkum}, \bibfnamefont {P.~and;~Schive}}, \bibinfo {author} {\bibfnamefont
  {T.}~\bibnamefont {Chiueh}},\ and\ \bibinfo {author} {\bibfnamefont
  {T.}~\bibnamefont {Broadhurst}},\ }\bibfield  {title} {\bibinfo {title} {On
  the random motion of nuclear objects in a fuzzy dark matter halo},\ }\href
  {https://doi.org/https://doi.org/10.3847/1538-4357/ac043f} {\bibfield
  {journal} {\bibinfo  {journal} {Astrophys. J.}\ }\textbf {\bibinfo {volume}
  {916}},\ \bibinfo {pages} {16} (\bibinfo {year} {2021})}\BibitemShut
  {NoStop}%
\bibitem [{\citenamefont {Church}\ \emph {et~al.}(2019)\citenamefont {Church},
  \citenamefont {Mocz},\ and\ \citenamefont {Ostriker}}]{Church:2019}%
  \BibitemOpen
  \bibfield  {author} {\bibinfo {author} {\bibfnamefont {B.~V.}\ \bibnamefont
  {Church}}, \bibinfo {author} {\bibfnamefont {P.}~\bibnamefont {Mocz}},\ and\
  \bibinfo {author} {\bibfnamefont {J.~P.}\ \bibnamefont {Ostriker}},\
  }\bibfield  {title} {\bibinfo {title} {Heating of milky way disc stars by
  dark matter fluctuations in cold dark matter and fuzzy dark matter
  paradigms.},\ }\href {https://doi.org/https://doi.org/10.1093/mnras/stz534}
  {\bibfield  {journal} {\bibinfo  {journal} {Mon. Not. R. Astron. Soc.}\
  }\textbf {\bibinfo {volume} {485}},\ \bibinfo {pages} {2861} (\bibinfo {year}
  {2019})}\BibitemShut {NoStop}%
\bibitem [{\citenamefont {Madau}\ and\ \citenamefont
  {Haardt}(2015)}]{Madau:2015}%
  \BibitemOpen
  \bibfield  {author} {\bibinfo {author} {\bibfnamefont {P.}~\bibnamefont
  {Madau}}\ and\ \bibinfo {author} {\bibfnamefont {F.}~\bibnamefont {Haardt}},\
  }\bibfield  {title} {\bibinfo {title} {Cosmic reionization after planck:
  Could quasars do it all?},\ }\href
  {https://doi.org/https://doi.org/10.1088/2041-8205/813/1/L8} {\bibfield
  {journal} {\bibinfo  {journal} {Astronomical. J. Lett.}\ }\textbf {\bibinfo
  {volume} {813}},\ \bibinfo {pages} {6} (\bibinfo {year} {2015})}\BibitemShut
  {NoStop}%
\bibitem [{\citenamefont {Padmanabhan}\ and\ \citenamefont
  {Loeb}(2021)}]{Padmanabhan:2021}%
  \BibitemOpen
  \bibfield  {author} {\bibinfo {author} {\bibfnamefont {H.}~\bibnamefont
  {Padmanabhan}}\ and\ \bibinfo {author} {\bibfnamefont {A.}~\bibnamefont
  {Loeb}},\ }\bibfield  {title} {\bibinfo {title} {Distinguishing agn from
  starbursts as the origin of double-peaked lyman-alpha emitters in the
  reionization era.},\ }\href
  {https://doi.org/https://doi.org/10.1051/0004-6361/202040107} {\bibfield
  {journal} {\bibinfo  {journal} {Astron. Astrophys.}\ }\textbf {\bibinfo
  {volume} {646}},\ \bibinfo {pages} {4} (\bibinfo {year} {2021})}\BibitemShut
  {NoStop}%
\bibitem [{\citenamefont {Hu}\ \emph {et~al.}(2016)\citenamefont {Hu},
  \citenamefont {Cowie}, \citenamefont {Songaila}, \citenamefont {Barger},
  \citenamefont {Rosenwasser},\ and\ \citenamefont {Wold}}]{Hu:2016}%
  \BibitemOpen
  \bibfield  {author} {\bibinfo {author} {\bibfnamefont {E.~M.}\ \bibnamefont
  {Hu}}, \bibinfo {author} {\bibfnamefont {L.~L.}\ \bibnamefont {Cowie}},
  \bibinfo {author} {\bibfnamefont {A.}~\bibnamefont {Songaila}}, \bibinfo
  {author} {\bibfnamefont {A.~J.}\ \bibnamefont {Barger}}, \bibinfo {author}
  {\bibfnamefont {B.}~\bibnamefont {Rosenwasser}},\ and\ \bibinfo {author}
  {\bibfnamefont {I.~G.~B.}\ \bibnamefont {Wold}},\ }\bibfield  {title}
  {\bibinfo {title} {An ultraluminous ly$\alpha$ emitter with a blue wing at z
  = 6.6.},\ }\href {https://doi.org/https://doi.org/10.3847/2041-8205/825/1/L7}
  {\bibfield  {journal} {\bibinfo  {journal} {Astronomical. J. Lett.}\ }\textbf
  {\bibinfo {volume} {825}},\ \bibinfo {pages} {5} (\bibinfo {year}
  {2016})}\BibitemShut {NoStop}%
\bibitem [{\citenamefont {Bosman}\ \emph {et~al.}(2020)\citenamefont {Bosman},
  \citenamefont {Kakiichi}, \citenamefont {Meyer}, \citenamefont {Gronke},
  \citenamefont {Laporte},\ and\ \citenamefont {Ellis}}]{Bosman:2020}%
  \BibitemOpen
  \bibfield  {author} {\bibinfo {author} {\bibfnamefont {S.~E.~I.}\
  \bibnamefont {Bosman}}, \bibinfo {author} {\bibfnamefont {K.}~\bibnamefont
  {Kakiichi}}, \bibinfo {author} {\bibfnamefont {R.~A.}\ \bibnamefont {Meyer}},
  \bibinfo {author} {\bibfnamefont {M.}~\bibnamefont {Gronke}}, \bibinfo
  {author} {\bibfnamefont {N.}~\bibnamefont {Laporte}},\ and\ \bibinfo {author}
  {\bibfnamefont {R.~S.}\ \bibnamefont {Ellis}},\ }\bibfield  {title} {\bibinfo
  {title} {Three ly$\alpha$ emitting galaxies within a quasar proximity zone at
  z $\simeq$ 5.8.},\ }\href
  {https://doi.org/https://doi.org/10.3847/1538-4357/ab85cd} {\bibfield
  {journal} {\bibinfo  {journal} {Astronomical. J.}\ }\textbf {\bibinfo
  {volume} {896}},\ \bibinfo {pages} {17} (\bibinfo {year} {2020})}\BibitemShut
  {NoStop}%
\bibitem [{\citenamefont {Gronke}\ \emph {et~al.}(2020)\citenamefont {Gronke},
  \citenamefont {Ocvirk}, \citenamefont {Mason}, \citenamefont {Matthee},
  \citenamefont {Bosman}, \citenamefont {Sorce}, \citenamefont {Lewis},
  \citenamefont {Ahn}, \citenamefont {Aubert}, \citenamefont {Dawoodbhoy},
  \citenamefont {Iliev}, \citenamefont {Shapiro},\ and\ \citenamefont
  {Yepes}}]{Gronke:2020}%
  \BibitemOpen
  \bibfield  {author} {\bibinfo {author} {\bibfnamefont {M.}~\bibnamefont
  {Gronke}}, \bibinfo {author} {\bibfnamefont {P.}~\bibnamefont {Ocvirk}},
  \bibinfo {author} {\bibfnamefont {C.}~\bibnamefont {Mason}}, \bibinfo
  {author} {\bibfnamefont {J.}~\bibnamefont {Matthee}}, \bibinfo {author}
  {\bibfnamefont {S.~I.}\ \bibnamefont {Bosman}}, \bibinfo {author}
  {\bibfnamefont {J.~G.}\ \bibnamefont {Sorce}}, \bibinfo {author}
  {\bibfnamefont {J.}~\bibnamefont {Lewis}}, \bibinfo {author} {\bibfnamefont
  {K.}~\bibnamefont {Ahn}}, \bibinfo {author} {\bibfnamefont {D.}~\bibnamefont
  {Aubert}}, \bibinfo {author} {\bibfnamefont {T.}~\bibnamefont {Dawoodbhoy}},
  \bibinfo {author} {\bibfnamefont {I.~T.}\ \bibnamefont {Iliev}}, \bibinfo
  {author} {\bibfnamefont {P.~R.}\ \bibnamefont {Shapiro}},\ and\ \bibinfo
  {author} {\bibfnamefont {G.}~\bibnamefont {Yepes}},\ }\bibfield  {title}
  {\bibinfo {title} {Lyman-alpha transmission properties of the intergalactic
  medium in the codaii simulation.},\ }\bibfield  {journal} {\bibinfo
  {journal} {Arxiv}\ }\href
  {https://doi.org/https://arxiv.org/pdf/2004.14496.pdf}
  {https://arxiv.org/pdf/2004.14496.pdf} (\bibinfo {year} {2020})\BibitemShut
  {NoStop}%
\bibitem [{\citenamefont {Becker}\ \emph {et~al.}(2015)\citenamefont {Becker},
  \citenamefont {Bolton}, \citenamefont {Madau}, \citenamefont {Pettini},
  \citenamefont {Ryan-Weber},\ and\ \citenamefont {Venemans}}]{Becker:2015}%
  \BibitemOpen
  \bibfield  {author} {\bibinfo {author} {\bibfnamefont {G.~D.}\ \bibnamefont
  {Becker}}, \bibinfo {author} {\bibfnamefont {J.~S.}\ \bibnamefont {Bolton}},
  \bibinfo {author} {\bibfnamefont {P.}~\bibnamefont {Madau}}, \bibinfo
  {author} {\bibfnamefont {M.}~\bibnamefont {Pettini}}, \bibinfo {author}
  {\bibfnamefont {E.~V.}\ \bibnamefont {Ryan-Weber}},\ and\ \bibinfo {author}
  {\bibfnamefont {B.~P.}\ \bibnamefont {Venemans}},\ }\bibfield  {title}
  {\bibinfo {title} {Evidence of patchy hydrogen reionization from an extreme
  ly$\alpha$ trough below redshift six},\ }\href
  {https://doi.org/https://doi.org/10.1093/mnras/stu2646} {\bibfield  {journal}
  {\bibinfo  {journal} {Mon. Not. R. Astron. Soc.}\ }\textbf {\bibinfo {volume}
  {447}},\ \bibinfo {pages} {3402} (\bibinfo {year} {2015})}\BibitemShut
  {NoStop}%
\bibitem [{\citenamefont {Gangolli}\ \emph {et~al.}(2021)\citenamefont
  {Gangolli}, \citenamefont {D'Aloisio}, \citenamefont {Nasir},\ and\
  \citenamefont {Zheng}}]{Gangolli:2021}%
  \BibitemOpen
  \bibfield  {author} {\bibinfo {author} {\bibfnamefont {N.}~\bibnamefont
  {Gangolli}}, \bibinfo {author} {\bibfnamefont {A.}~\bibnamefont {D'Aloisio}},
  \bibinfo {author} {\bibfnamefont {F.}~\bibnamefont {Nasir}},\ and\ \bibinfo
  {author} {\bibfnamefont {Z.}~\bibnamefont {Zheng}},\ }\bibfield  {title}
  {\bibinfo {title} {Constraining reionization in progress at z = 5.7 with
  lyman-$\alpha$ emitters: voids, peaks, and cosmic variance.},\ }\href
  {https://doi.org/https://doi.org/10.1093/mnras/staa3843} {\bibfield
  {journal} {\bibinfo  {journal} {Mon. Not. R. Astron. Soc.}\ }\textbf
  {\bibinfo {volume} {501}},\ \bibinfo {pages} {5294} (\bibinfo {year}
  {2021})}\BibitemShut {NoStop}%
\bibitem [{\citenamefont {Dalal}\ and\ \citenamefont
  {Kravtsov}(2022)}]{Dalal:2022}%
  \BibitemOpen
  \bibfield  {author} {\bibinfo {author} {\bibfnamefont {N.}~\bibnamefont
  {Dalal}}\ and\ \bibinfo {author} {\bibfnamefont {A.}~\bibnamefont
  {Kravtsov}},\ }\bibfield  {title} {\bibinfo {title} {Not so fuzzy: excluding
  fdm with sizes and stellar kinematics of ultra-faint dwarf galaxies.},\
  }\bibfield  {journal} {\bibinfo  {journal} {ArXiv}\ }\href
  {https://doi.org/https://arxiv.org/pdf/2203.05750.pdf}
  {https://arxiv.org/pdf/2203.05750.pdf} (\bibinfo {year} {2022})\BibitemShut
  {NoStop}%
\bibitem [{\citenamefont {Pozo}\ \emph {et~al.}(2024)\citenamefont {Pozo},
  \citenamefont {Broadhurst}, \citenamefont {Smoot}, \citenamefont {Chiueh},
  \citenamefont {L.}, \citenamefont {Vogelsberger},\ and\ \citenamefont
  {Mocz}}]{Pozo:2024}%
  \BibitemOpen
  \bibfield  {author} {\bibinfo {author} {\bibfnamefont {A.}~\bibnamefont
  {Pozo}}, \bibinfo {author} {\bibfnamefont {T.}~\bibnamefont {Broadhurst}},
  \bibinfo {author} {\bibfnamefont {G.~F.}\ \bibnamefont {Smoot}}, \bibinfo
  {author} {\bibfnamefont {T.}~\bibnamefont {Chiueh}}, \bibinfo {author}
  {\bibfnamefont {N.}~\bibnamefont {L.}}, \bibinfo {author} {\bibfnamefont
  {M.}~\bibnamefont {Vogelsberger}},\ and\ \bibinfo {author} {\bibfnamefont
  {P.}~\bibnamefont {Mocz}},\ }\bibfield  {title} {\bibinfo {title} {Dwarf
  galaxies united by dark bosons.},\ }\bibfield  {journal} {\bibinfo  {journal}
  {Phys. Rev. D.}\ }\textbf {\bibinfo {volume} {109}},\ \href
  {https://doi.org/https://doi.org/10.1103/PhysRevD.109.083532}
  {https://doi.org/10.1103/PhysRevD.109.083532} (\bibinfo {year}
  {2024})\BibitemShut {NoStop}%
\bibitem [{\citenamefont {Amruth}\ \emph {et~al.}(2023)\citenamefont {Amruth},
  \citenamefont {Broadhurst}, \citenamefont {;~Lim}, \citenamefont {Oguri},
  \citenamefont {Smoot}, \citenamefont {Diego}, \citenamefont {Leung},
  \citenamefont {Emami}, \citenamefont {Li}, \citenamefont {Chiueh},
  \citenamefont {Schive}, \citenamefont {Yeung},\ and\ \citenamefont
  {Li}}]{Amruth:2023}%
  \BibitemOpen
  \bibfield  {author} {\bibinfo {author} {\bibfnamefont {A.}~\bibnamefont
  {Amruth}}, \bibinfo {author} {\bibfnamefont {T.}~\bibnamefont {Broadhurst}},
  \bibinfo {author} {\bibfnamefont {J.}~\bibnamefont {;~Lim}}, \bibinfo
  {author} {\bibfnamefont {M.}~\bibnamefont {Oguri}}, \bibinfo {author}
  {\bibfnamefont {G.~F.}\ \bibnamefont {Smoot}}, \bibinfo {author}
  {\bibfnamefont {J.~M.}\ \bibnamefont {Diego}}, \bibinfo {author}
  {\bibfnamefont {E.}~\bibnamefont {Leung}}, \bibinfo {author} {\bibfnamefont
  {R.}~\bibnamefont {Emami}}, \bibinfo {author} {\bibfnamefont
  {J.}~\bibnamefont {Li}}, \bibinfo {author} {\bibfnamefont {T.}~\bibnamefont
  {Chiueh}}, \bibinfo {author} {\bibfnamefont {H.-Y.}\ \bibnamefont {Schive}},
  \bibinfo {author} {\bibfnamefont {M.~C.~H.}\ \bibnamefont {Yeung}},\ and\
  \bibinfo {author} {\bibfnamefont {S.~K.}\ \bibnamefont {Li}},\ }\bibfield
  {title} {\bibinfo {title} {Einstein rings modulated by wavelike dark matter
  from anomalies in gravitationally lensed images.},\ }\href
  {https://doi.org/https://doi.org/10.1038/s41550-023-01943-9} {\bibfield
  {journal} {\bibinfo  {journal} {Nature Astro}\ }\textbf {\bibinfo {volume}
  {7}},\ \bibinfo {pages} {736} (\bibinfo {year} {2023})}\BibitemShut {NoStop}%
\bibitem [{\citenamefont {de~Martino}\ \emph {et~al.}(2017)\citenamefont
  {de~Martino}, \citenamefont {Broadhurst}, \citenamefont {Tye}, \citenamefont
  {Chiueh}, \citenamefont {Schive},\ and\ \citenamefont
  {Lazkoz}}]{deMartino:2017}%
  \BibitemOpen
  \bibfield  {author} {\bibinfo {author} {\bibfnamefont {I.}~\bibnamefont
  {de~Martino}}, \bibinfo {author} {\bibfnamefont {T.}~\bibnamefont
  {Broadhurst}}, \bibinfo {author} {\bibfnamefont {S.~H.~H.}\ \bibnamefont
  {Tye}}, \bibinfo {author} {\bibfnamefont {T.}~\bibnamefont {Chiueh}},
  \bibinfo {author} {\bibfnamefont {H.-Y.}\ \bibnamefont {Schive}},\ and\
  \bibinfo {author} {\bibfnamefont {R.}~\bibnamefont {Lazkoz}},\ }\bibfield
  {title} {\bibinfo {title} {Recognising axionic dark matter by compton and
  de-broglie scale modulation of pulsar timing.},\ }\bibfield  {journal}
  {\bibinfo  {journal} {Phys. Rev. Lett.}\ }\textbf {\bibinfo {volume} {119}},\
  \href {https://doi.org/https://doi.org/10.1103/PhysRevLett.119.221103}
  {https://doi.org/10.1103/PhysRevLett.119.221103} (\bibinfo {year}
  {2017})\BibitemShut {NoStop}%
\bibitem [{\citenamefont {Luu}\ \emph {et~al.}(2024)\citenamefont {Luu},
  \citenamefont {Liu}, \citenamefont {Ren}, \citenamefont {Broadhurst},
  \citenamefont {Yang}, \citenamefont {Wang},\ and\ \citenamefont
  {Xie}}]{Luu:2024}%
  \BibitemOpen
  \bibfield  {author} {\bibinfo {author} {\bibfnamefont {H.~N.}\ \bibnamefont
  {Luu}}, \bibinfo {author} {\bibfnamefont {T.}~\bibnamefont {Liu}}, \bibinfo
  {author} {\bibfnamefont {J.}~\bibnamefont {Ren}}, \bibinfo {author}
  {\bibfnamefont {T.}~\bibnamefont {Broadhurst}}, \bibinfo {author}
  {\bibfnamefont {R.}~\bibnamefont {Yang}}, \bibinfo {author} {\bibfnamefont
  {J.-S.}\ \bibnamefont {Wang}},\ and\ \bibinfo {author} {\bibfnamefont
  {Z.}~\bibnamefont {Xie}},\ }\bibfield  {title} {\bibinfo {title} {Stochastic
  wave dark matter with fermi-lat $\gamma$-ray pulsar timing array},\ }\href
  {https://doi.org/https://doi.org/10.3847/2041-8213/ad2ae2} {\bibfield
  {journal} {\bibinfo  {journal} {Astronomical. J. Lett.}\ }\textbf {\bibinfo
  {volume} {963}},\ \bibinfo {pages} {8} (\bibinfo {year} {2024})}\BibitemShut
  {NoStop}%
\bibitem [{\citenamefont {Roy}(2020)}]{Roy:2020}%
  \BibitemOpen
  \bibfield  {author} {\bibinfo {author} {\bibfnamefont {V.}~\bibnamefont
  {Roy}},\ }\bibfield  {title} {\bibinfo {title} {Convergence diagnostics for
  markov chain monte carlo.},\ }\bibfield  {journal} {\bibinfo  {journal}
  {Annual Rev. of Statis. and Its App.}\ }\textbf {\bibinfo {volume} {7}},\
  \href
  {https://doi.org/https://doi.org/10.1146/annurev-statistics-031219-041300}
  {https://doi.org/10.1146/annurev-statistics-031219-041300} (\bibinfo {year}
  {2020})\BibitemShut {NoStop}%
\bibitem [{\citenamefont {Cicuéndez}\ and\ \citenamefont
  {Battaglia}(2018)}]{Cicuendez:2018}%
  \BibitemOpen
  \bibfield  {author} {\bibinfo {author} {\bibfnamefont {L.}~\bibnamefont
  {Cicuéndez}}\ and\ \bibinfo {author} {\bibfnamefont {G.}~\bibnamefont
  {Battaglia}},\ }\bibfield  {title} {\bibinfo {title} {Appearances can be
  deceiving: clear signs of accretion in the seemingly ordinary sextans dsph},\
  }\href {https://doi.org/https://doi.org/10.1093/mnras/sty1748} {\bibfield
  {journal} {\bibinfo  {journal} {Mon. Not. R. Astron. Soc.}\ }\textbf
  {\bibinfo {volume} {480}},\ \bibinfo {pages} {251} (\bibinfo {year}
  {2018})}\BibitemShut {NoStop}%
\bibitem [{\citenamefont {Zaggia}\ \emph {et~al.}(2011)\citenamefont {Zaggia},
  \citenamefont {Held}, \citenamefont {Sommariva}, \citenamefont {Momany},
  \citenamefont {Saviane},\ and\ \citenamefont {Rizzi}}]{Zaggia:2011}%
  \BibitemOpen
  \bibfield  {author} {\bibinfo {author} {\bibfnamefont {S.}~\bibnamefont
  {Zaggia}}, \bibinfo {author} {\bibfnamefont {E.~V.}\ \bibnamefont {Held}},
  \bibinfo {author} {\bibfnamefont {V.}~\bibnamefont {Sommariva}}, \bibinfo
  {author} {\bibfnamefont {Y.}~\bibnamefont {Momany}}, \bibinfo {author}
  {\bibfnamefont {I.}~\bibnamefont {Saviane}},\ and\ \bibinfo {author}
  {\bibfnamefont {L.}~\bibnamefont {Rizzi}},\ }\bibfield  {title} {\bibinfo
  {title} {Phoenix dwarf galaxy stellar kinematics},\ }\href
  {https://doi.org/https://doi.org/10.1051/eas/1148049} {\bibfield  {journal}
  {\bibinfo  {journal} {EAS Publications Series}\ }\textbf {\bibinfo {volume}
  {48}},\ \bibinfo {pages} {215} (\bibinfo {year} {2011})}\BibitemShut
  {NoStop}%
\bibitem [{\citenamefont {Koch}\ \emph {et~al.}(2007)\citenamefont {Koch},
  \citenamefont {Wilkinson}, \citenamefont {Kleyna}, \citenamefont {Gilmore},
  \citenamefont {Grebel}, \citenamefont {Mackey}, \citenamefont {Evans},\ and\
  \citenamefont {Wyse}}]{Koch:2007}%
  \BibitemOpen
  \bibfield  {author} {\bibinfo {author} {\bibfnamefont {A.}~\bibnamefont
  {Koch}}, \bibinfo {author} {\bibfnamefont {M.~I.}\ \bibnamefont {Wilkinson}},
  \bibinfo {author} {\bibfnamefont {J.~T.}\ \bibnamefont {Kleyna}}, \bibinfo
  {author} {\bibfnamefont {G.~F.}\ \bibnamefont {Gilmore}}, \bibinfo {author}
  {\bibfnamefont {E.~K.}\ \bibnamefont {Grebel}}, \bibinfo {author}
  {\bibfnamefont {A.~D.}\ \bibnamefont {Mackey}}, \bibinfo {author}
  {\bibfnamefont {N.~W.}\ \bibnamefont {Evans}},\ and\ \bibinfo {author}
  {\bibfnamefont {R.~F.~G.}\ \bibnamefont {Wyse}},\ }\bibfield  {title}
  {\bibinfo {title} {Stellar kinematics and metallicities in the leo i dwarf
  spheroidal galaxy-wide-field implications for galactic evolution},\ }\href
  {https://doi.org/https://doi.org/10.1086/510879} {\bibfield  {journal}
  {\bibinfo  {journal} {Astrophys. J.}\ }\textbf {\bibinfo {volume} {657}},\
  \bibinfo {pages} {241} (\bibinfo {year} {2007})}\BibitemShut {NoStop}%
\bibitem [{\citenamefont {Łokas}\ and\ \citenamefont
  {Mamon}(2005)}]{Lokas:2005}%
  \BibitemOpen
  \bibfield  {author} {\bibinfo {author} {\bibfnamefont {E.~L.}\ \bibnamefont
  {Łokas}}\ and\ \bibinfo {author} {\bibfnamefont {F.}~\bibnamefont {Mamon},
  \bibfnamefont {Gary A. search by orcid ;~Prada}},\ }\bibfield  {title}
  {\bibinfo {title} {Dark matter distribution in the draco dwarf from velocity
  moments},\ }\href {https://doi.org/https://doi.org/
  10.1111/j.1365-2966.2005.09497.x} {\bibfield  {journal} {\bibinfo  {journal}
  {Mon. Not. R. Astron. Soc.}\ }\textbf {\bibinfo {volume} {363}},\ \bibinfo
  {pages} {918} (\bibinfo {year} {2005})}\BibitemShut {NoStop}%
\bibitem [{\citenamefont {Walker}\ \emph {et~al.}(2009)\citenamefont {Walker},
  \citenamefont {Mateo}, \citenamefont {Olszewski}, \citenamefont
  {Peñarrubia}, \citenamefont {Evans},\ and\ \citenamefont
  {Gilmore}}]{Walker:2009}%
  \BibitemOpen
  \bibfield  {author} {\bibinfo {author} {\bibfnamefont {M.~G.}\ \bibnamefont
  {Walker}}, \bibinfo {author} {\bibfnamefont {M.}~\bibnamefont {Mateo}},
  \bibinfo {author} {\bibfnamefont {E.~W.}\ \bibnamefont {Olszewski}}, \bibinfo
  {author} {\bibfnamefont {J.}~\bibnamefont {Peñarrubia}}, \bibinfo {author}
  {\bibfnamefont {N.~W.}\ \bibnamefont {Evans}},\ and\ \bibinfo {author}
  {\bibfnamefont {G.}~\bibnamefont {Gilmore}},\ }\bibfield  {title} {\bibinfo
  {title} {A universal mass profile for dwarf spheroidal galaxies?},\ }\href
  {https://doi.org/https://doi.org/10.1088/0004-637X/704/2/1274} {\bibfield
  {journal} {\bibinfo  {journal} {Astrophys. J.}\ }\textbf {\bibinfo {volume}
  {704}},\ \bibinfo {pages} {1274} (\bibinfo {year} {2009})}\BibitemShut
  {NoStop}%
\bibitem [{\citenamefont {Fritz}\ \emph {et~al.}(2018)\citenamefont {Fritz},
  \citenamefont {Pawlowski}, \citenamefont {Kallivayalil}, \citenamefont
  {van~der Marel}, \citenamefont {Sohn}, \citenamefont {Brook},\ and\
  \citenamefont {Besla}}]{Fritz:2018}%
  \BibitemOpen
  \bibfield  {author} {\bibinfo {author} {\bibfnamefont {G.}~\bibnamefont
  {Fritz}, \bibfnamefont {T.~K.and~Battaglia}}, \bibinfo {author}
  {\bibfnamefont {M.~S.}\ \bibnamefont {Pawlowski}}, \bibinfo {author}
  {\bibfnamefont {N.}~\bibnamefont {Kallivayalil}}, \bibinfo {author}
  {\bibfnamefont {R.}~\bibnamefont {van~der Marel}}, \bibinfo {author}
  {\bibfnamefont {S.~T.}\ \bibnamefont {Sohn}}, \bibinfo {author}
  {\bibfnamefont {C.}~\bibnamefont {Brook}},\ and\ \bibinfo {author}
  {\bibfnamefont {G.}~\bibnamefont {Besla}},\ }\bibfield  {title} {\bibinfo
  {title} {Gaia dr2 proper motions of dwarf galaxies within 420 kpc. orbits,
  milky way mass, tidal influences, planar alignments, and group infall.},\
  }\href {https://doi.org/https://doi.org/10.1051/0004-6361/201833343}
  {\bibfield  {journal} {\bibinfo  {journal} {Astron. Astrophys.}\ }\textbf
  {\bibinfo {volume} {619}},\ \bibinfo {pages} {18} (\bibinfo {year}
  {2018})}\BibitemShut {NoStop}%
\bibitem [{\citenamefont {McConnachie}\ and\ \citenamefont
  {Venn}(2020)}]{McConnachie:2020}%
  \BibitemOpen
  \bibfield  {author} {\bibinfo {author} {\bibfnamefont {A.~W.}\ \bibnamefont
  {McConnachie}}\ and\ \bibinfo {author} {\bibfnamefont {K.~A.}\ \bibnamefont
  {Venn}},\ }\bibfield  {title} {\bibinfo {title} {Revised and new proper
  motions for confirmed and candidate milky way dwarf galaxies.},\ }\bibfield
  {journal} {\bibinfo  {journal} {Astronomical. J.}\ }\textbf {\bibinfo
  {volume} {160}},\ \href
  {https://doi.org/https://doi.org/10.3847/1538-3881/aba4ab}
  {https://doi.org/10.3847/1538-3881/aba4ab} (\bibinfo {year}
  {2020})\BibitemShut {NoStop}%
\bibitem [{\citenamefont {Frinchaboy}\ \emph {et~al.}(2012)\citenamefont
  {Frinchaboy}, \citenamefont {Majewski}, \citenamefont {Muñoz}, \citenamefont
  {Law}, \citenamefont {Łokas}, \citenamefont {Kunkel}, \citenamefont
  {Patterson},\ and\ \citenamefont {Johnston}}]{Frinchaboy:2012}%
  \BibitemOpen
  \bibfield  {author} {\bibinfo {author} {\bibfnamefont {P.~M.}\ \bibnamefont
  {Frinchaboy}}, \bibinfo {author} {\bibfnamefont {S.~R.}\ \bibnamefont
  {Majewski}}, \bibinfo {author} {\bibfnamefont {R.~R.}\ \bibnamefont
  {Muñoz}}, \bibinfo {author} {\bibfnamefont {D.~R.}\ \bibnamefont {Law}},
  \bibinfo {author} {\bibfnamefont {E.~L.}\ \bibnamefont {Łokas}}, \bibinfo
  {author} {\bibfnamefont {W.~E.}\ \bibnamefont {Kunkel}}, \bibinfo {author}
  {\bibfnamefont {R.}~\bibnamefont {Patterson}},\ and\ \bibinfo {author}
  {\bibfnamefont {K.~V.}\ \bibnamefont {Johnston}},\ }\bibfield  {title}
  {\bibinfo {title} {A 2mass all-sky view of the sagittarius dwarf galaxy. vii.
  kinematics of the main body of the sagittarius},\ }\href
  {https://doi.org/https://doi.org/10.1088/0004-637X/756/1/74} {\bibfield
  {journal} {\bibinfo  {journal} {Astrophys. J.}\ }\textbf {\bibinfo {volume}
  {756}},\ \bibinfo {pages} {19} (\bibinfo {year} {2012})}\BibitemShut
  {NoStop}%
\bibitem [{\citenamefont {Coleman}\ \emph {et~al.}(2007)\citenamefont
  {Coleman}, \citenamefont {Rix}, \citenamefont {Grebel},\ and\ \citenamefont
  {Koch}}]{Coleman:2007}%
  \BibitemOpen
  \bibfield  {author} {\bibinfo {author} {\bibfnamefont {K.}~\bibnamefont
  {Coleman}, \bibfnamefont {M.~G.and~Jordi}}, \bibinfo {author} {\bibfnamefont
  {H.-W.}\ \bibnamefont {Rix}}, \bibinfo {author} {\bibfnamefont {E.~K.}\
  \bibnamefont {Grebel}},\ and\ \bibinfo {author} {\bibfnamefont
  {A.}~\bibnamefont {Koch}},\ }\bibfield  {title} {\bibinfo {title} {A
  wide-field view of leo ii: A structural analysis using the sloan digital sky
  survey.},\ }\href {https://doi.org/https://doi.org/10.1086/522229} {\bibfield
   {journal} {\bibinfo  {journal} {Astronomical. J.}\ }\textbf {\bibinfo
  {volume} {134}},\ \bibinfo {pages} {1938} (\bibinfo {year}
  {2007})}\BibitemShut {NoStop}%
\bibitem [{\citenamefont {Battaglia}\ \emph
  {et~al.}(2012{\natexlab{b}})\citenamefont {Battaglia}, \citenamefont
  {Rejkuba}, \citenamefont {Tolstoy}, \citenamefont {Irwin},\ and\
  \citenamefont {Beccari}}]{Battaglia:2012}%
  \BibitemOpen
  \bibfield  {author} {\bibinfo {author} {\bibfnamefont {G.}~\bibnamefont
  {Battaglia}}, \bibinfo {author} {\bibfnamefont {M.}~\bibnamefont {Rejkuba}},
  \bibinfo {author} {\bibfnamefont {E.}~\bibnamefont {Tolstoy}}, \bibinfo
  {author} {\bibfnamefont {M.~J.}\ \bibnamefont {Irwin}},\ and\ \bibinfo
  {author} {\bibfnamefont {G.~A.}\ \bibnamefont {Beccari}},\ }\bibfield
  {title} {\bibinfo {title} {A wide-area view of the phoenix dwarf galaxy from
  very large telescope/fors imaging.},\ }\href
  {https://doi.org/https://doi.org/10.1111/j.1365-2966.2012.21286.x} {\bibfield
   {journal} {\bibinfo  {journal} {Mon. Not. R. Astron. Soc.}\ }\textbf
  {\bibinfo {volume} {424}},\ \bibinfo {pages} {1113} (\bibinfo {year}
  {2012}{\natexlab{b}})}\BibitemShut {NoStop}%
\end{thebibliography}%

\end{document}